\documentclass[aps,prb,twocolumn,superscriptaddress,floatfix,longbibliography]{revtex4-2}

\usepackage{amsmath,amssymb}
\usepackage{bm}
\usepackage{graphicx}
\usepackage{dcolumn}
\usepackage[dvipsnames]{xcolor}
\usepackage[colorlinks=true,linkcolor=red,citecolor=blue,urlcolor=magenta]{hyperref}

\graphicspath{{figures/}}

\begin{document}

\title{The First Magic Angle Beyond the Chiral Limit in Twisted Bilayer Graphene}

\author{Leonardo A. Navarro-Labastida}
\affiliation{Depto.\ de Sistemas Complejos, Instituto de F\'isica,\\ Universidad Nacional Aut\'onoma de M\'exico (UNAM). Apdo.\ Postal 20-364, 01000 M\'exico D.F., M\'exico}
\affiliation{Arts and Sciences, NYU Shanghai, Shanghai 200124, China}
\affiliation{NYU-ECNU Institute of Physics at NYU Shanghai, Shanghai 200062, China}

\author{Pierre A. Pantale\'on}
\email{pierre.pantaleon@hainanu.edu.cn}
\affiliation{Center for Theoretical Physics, Hainan University, Haikou 570228, China}
\affiliation{School of Physics and Optoelectronic Engineering, Hainan University, Haikou 570228, China}
\affiliation{IMDEA Nanoscience, Faraday 9, 28049 Madrid, Spain}

\author{Francisco Guinea}
\affiliation{IMDEA Nanoscience, Faraday 9, 28049 Madrid, Spain}
\affiliation{Donostia International Physics Center, Paseo Manuel de Lardiz\'abal 4, 20018 San Sebasti\'an, Spain}

\author{Gerardo G. Naumis}
\email{naumis@fisica.unam.mx}
\affiliation{Depto.\ de Sistemas Complejos, Instituto de F\'isica,\\ Universidad Nacional Aut\'onoma de M\'exico (UNAM). Apdo.\ Postal 20-364, 01000 M\'exico D.F., M\'exico}

\begin{abstract}
We develop a squared-Hamiltonian description of twisted bilayer graphene beyond the chiral limit to explain why the first magic angle remains robust under lattice relaxation, while higher-order magic angles are strongly destabilized. Starting from the non-chiral Bistritzer--MacDonald model with finite same-sublattice tunneling, we show that lattice relaxation reshapes the effective confinement landscape rather than acting as a simple perturbation of the chiral theory. A central result is that the realistic relaxation-renormalized tunneling ratio lies close to a special confinement point where the oscillatory part of the symmetric confinement potential nearly cancels. This places realistic twisted bilayer graphene near a nearly uniform confinement regime. At the same time, finite same-sublattice tunneling activates an additional inter-sublattice current-like channel that competes with the chiral orbital channel. The first magic angle survives because these confinement and current-like contributions remain balanced, whereas higher-order magic angles lose this balance through stronger remote-band hybridization and enhanced real-space localization around AA regions. Our results provide a single-particle mechanism for the breakdown of the chiral magic-angle hierarchy and clarify why the experimentally relevant first magic angle remains the most stable remnant of the chiral flat-band structure.
\end{abstract}

\maketitle

\section{Introduction}

Twisted van der Waals (vdW) materials provide a versatile platform for controlling electronic structure through geometry. A small relative rotation between adjacent layers produces a moir\'e superlattice with a period much larger than the atomic lattice constant, strongly modifying the low-energy spectrum and enhancing the role of interactions, topology, and quantum interference effects~\cite{Andrei2020,Balents2020,MacDonald2021,Bistritzer2011a,Qin2021,Carr2020,Kennes2021}. Among these systems, twisted bilayer graphene (TBG) remains the paradigmatic example. Near the first magic angle, approximately $1.1^\circ$, the electronic bandwidth close to charge neutrality is strongly suppressed, producing narrow bands in which Coulomb interactions become comparable to, or larger than, the kinetic energy scale~\cite{Bistritzer2011a}. This regime has enabled correlated insulating states, superconductivity, orbital magnetism, and topological phases~\cite{Balents2020,Cao2018a,Cao2018b,Sharpe2019,Serlin2020,SolisFernandez2022}.

The broader relevance of moir\'e materials follows from the same geometric mechanism. By enlarging the unit cell, the moir\'e pattern rescales the characteristic kinetic, interaction, and magnetic-field energy scales, allowing regimes that are difficult to access in ordinary two-dimensional crystals~\cite{Andrei2020,Balents2020,Bistritzer2011a,Qin2021}. This tunability has motivated extensive studies not only of TBG, but also of twisted transition-metal dichalcogenides, multilayer graphene systems, heterobilayers, hybrid moir\'e structures, and one-dimensional moir\'e systems such as double-walled carbon nanotubes~\cite{Wang2020,Regan2020,Tang2020,Koshino2015a,Koshino2015b,Zhao2022,Sun2025,Tan2026}. These platforms exhibit a wide range of correlated and topological phenomena, including generalized Wigner crystals, Chern insulators, fractional Chern insulators, heavy-fermion-like behavior, magnetic textures, and superconducting phases~\cite{Kennes2021,Sharpe2019,Serlin2020,Regan2020,Tang2020,Xu2022,Escudero2026}. The same control over twist angle, pressure, electrostatic gating, strain, and external fields has also motivated proposals for quantum simulation and quantum-information applications based on moir\'e flat bands and interaction-driven topological states~\cite{Balents2020,Kennes2021,NavarroQIP2025,NavarroRMF2022}.

A central theoretical description of TBG is the continuum model of Bistritzer and MacDonald, in which two rotated Dirac Hamiltonians are coupled by spatially modulated interlayer tunneling terms~\cite{Bistritzer2011a}. This model captures the emergence of narrow bands at a sequence of magic angles. A particularly important idealization is the chiral limit, obtained by suppressing the same-sublattice tunneling amplitude while keeping the opposite-sublattice tunneling finite. In this limit, the Hamiltonian has an additional chiral symmetry and supports exactly flat bands at a discrete set of magic angles~\cite{Tarnopolsky2019a}. The chiral model has revealed a highly structured mathematical problem, including zero-mode wave functions, a hierarchy of magic angles, a characteristic scaling rule, and close analogies with Landau-level physics~\cite{Tarnopolsky2019a,Khalaf2019a}. These developments connect TBG to broader ideas in quantum geometry, generalized Landau-level descriptions of Chern bands, and effective gauge-field formulations of moir\'e flat bands~\cite{Liu2025,Mera2024,Ozawa2024,Montag2026}.

Realistic TBG, however, is not exactly chiral. Atomic relaxation changes the relative area of local stacking regions, reducing the effective tunneling through AA regions while enlarging the energetically favorable AB and BA domains~\cite{Nam2017,Carr2018,Koshino2018}. This effect is commonly incorporated in the continuum model through a finite ratio between same-sublattice and opposite-sublattice interlayer tunneling amplitudes. In addition, realistic systems may contain strain, particle-hole asymmetry, interlayer-coupling anisotropy, and other symmetry-breaking perturbations~\cite{Nam2017,Carr2018,Koshino2018}. These corrections modify the ideal chiral spectrum. In particular, numerical and analytical studies indicate that the higher-order chiral magic angles are strongly destabilized, whereas the first magic angle remains comparatively robust near experimentally relevant parameters~\cite{Tarnopolsky2019a,Koshino2018,Carr2019,Carr2020b}. Understanding why this first angle survives more effectively than the higher-order members of the chiral hierarchy remains an important theoretical question.

One useful route to this problem is provided by the squared-Hamiltonian formulation of the chiral model. In this representation, the original Dirac-like continuum Hamiltonian is mapped to an effective Schr\"odinger-like operator containing confinement terms, pseudo-magnetic gauge fields, and interlayer matrix elements. Previous work has related this structure to effective magnetic flux scaling, current-like interlayer contributions, and quantum-Hall-type descriptions in momentum or moir\'e space~\cite{Navarro2023,Navarro2022,Navarro2023RMF,Naumis2021,Navarro2025JPCS,Navarro2024,Navarro2026}. These ideas suggest that the loss of exact flatness in the non-chiral model should not be viewed only as a numerical shift of the band energies, but rather as a modification of the effective confinement and interlayer coupling structure inherited from the chiral problem.

In this work, we study this evolution from the chiral to the non-chiral Bistritzer--MacDonald model within an effective single-particle framework. We focus on how a finite same-sublattice tunneling ratio, motivated by lattice relaxation, modifies the squared Hamiltonian and the associated flat-band hierarchy. We show that finite same-sublattice tunneling renormalizes the diagonal confinement potential, introduces additional off-diagonal interlayer matrix elements, enhances real-space localization around AA regions, and increases hybridization between the central bands and remote bands. A key outcome is that the commonly used relaxation-renormalized value $\kappa\simeq0.7$ lies close to $\kappa=1/\sqrt{2}$, where the oscillatory part of the symmetric confinement in the squared Hamiltonian cancels. These effects suppress the higher-order chiral magic-angle structure while leaving the first magic angle comparatively robust. Rather than invoking many-body physics, the analysis identifies the single-particle mechanisms by which non-chiral tunneling modifies spectral squeezing, wave-function localization, and current-like interlayer channels in magic-angle TBG.

\section{Chiral and non-chiral TBG Hamiltonians}

The Bistritzer--MacDonald (BM) Hamiltonian provides the standard continuum description of the low-energy electronic structure of twisted bilayer graphene (TBG)~\cite{Bistritzer2011b}. It is constructed from the Dirac Hamiltonians of two rotated graphene layers coupled by spatially modulated interlayer tunneling terms. The dominant tunneling processes are parametrized by the amplitudes $w_{AA}$ and $w_{AB}$, which describe same-sublattice tunneling through local AA regions and opposite-sublattice tunneling through AB/BA regions, respectively.

Lattice relaxation plays an essential role in realistic TBG. Atomic reconstruction reduces the spatial extent of AA regions and enlarges the energetically favorable AB/BA domains, thereby suppressing the effective AA tunneling amplitude~\cite{Nam2017,Carr2018,Koshino2018}. The ideal chiral limit is obtained by setting $w_{AA}=0$ while keeping $w_{AB}$ finite. This limit, introduced by Tarnopolsky, Kruchkov, and Vishwanath, gives the Hamiltonian an exact chiral symmetry and supports exactly flat bands at a discrete set of magic angles~\cite{Tarnopolsky2019b}. It therefore provides the natural reference point for analyzing how finite same-sublattice tunneling modifies the magic-angle hierarchy.

We work in a single-valley continuum theory. In the sublattice basis, the four-component bi-spinor is written as
\begin{equation}
|\Phi\rangle=\big(\psi_1(\bm r),\psi_2(\bm r),\chi_1(\bm r),\chi_2(\bm r)\big)^T,
\end{equation}
where the indices $1,2$ label the graphene layers. The amplitudes $\psi_j(\bm r)$ and $\chi_j(\bm r)$ correspond to the continuum envelope functions associated with the two graphene sublattices. It is useful to introduce the layer spinors
\begin{equation}
\Psi=\begin{pmatrix}\psi_1(\bm r)\\ \psi_2(\bm r)\end{pmatrix},\qquad
\chi=\begin{pmatrix}\chi_1(\bm r)\\ \chi_2(\bm r)\end{pmatrix}.
\end{equation}
In this representation, $\Psi$ and $\chi$ define the two sublattice sectors, while the entries inside each spinor refer to the layer degree of freedom. The numerical calculations are performed in a Bloch plane-wave basis. For the $\Psi$ sector, we use
\begin{equation}
\begin{pmatrix}\psi_1(\bm r)\\ \psi_2(\bm r)\end{pmatrix}
=\sum_{m,n}\begin{pmatrix}a_{m,n}\\ b_{m,n}e^{i\bm q_1\cdot\bm r}\end{pmatrix}e^{i(\bm k+\bm K_{m,n})\cdot\bm r},
\end{equation}
where $a_{m,n}$ and $b_{m,n}$ are Fourier coefficients, $\bm K_{m,n}=m\bm b_1+n\bm b_2$ is a moir\'e reciprocal vector, and $\bm k$ is the crystal momentum in the moir\'e Brillouin zone. The same type of Bloch expansion is used for the $\chi$ sector. The additional phase in the second layer fixes the gauge convention used below for the interlayer coupling functions.

In the sublattice-block basis $(\Psi,\chi)$, the non-chiral BM Hamiltonian can be written as~\cite{Tarnopolsky2019b,Khalaf2019b,Ledwith2020,Popov2023a}
\begin{equation}
\mathcal H=\begin{pmatrix} M & D^\dagger\\ D & M\end{pmatrix}.
\label{eq:H}
\end{equation}
The operators $D$ and $D^\dagger$ contain the Dirac kinetic energy and the opposite-sublattice interlayer tunneling,
\begin{equation}
D(\bm r)=\begin{pmatrix} -i\bar\partial & \alpha U_1(\bm r)\\ \alpha U_1(-\bm r) & -i\bar\partial\end{pmatrix},
\end{equation}
and
\begin{equation}
D^\dagger(\bm r)=\begin{pmatrix} -i\partial & \alpha U_{-1}(\bm r)\\ \alpha U_{-1}(-\bm r) & -i\partial\end{pmatrix}.
\end{equation}
Here $\bar\partial=\partial_x+i\partial_y$ and $\partial=\partial_x-i\partial_y$. The functions $U_m(\bm r)$ describe the moir\'e modulation of the interlayer coupling,
\begin{equation}
U_m(\bm r)=\sum_{\nu=1}^{3}e^{im\phi(\nu-1)}e^{-i\bm q_\nu\cdot\bm r},
\end{equation}
where $\phi=2\pi/3$ and $m=0,\pm1$. With the conventions used here, these functions satisfy
\begin{equation}
U_1(\bm r)=-\bar\partial U_0(\bm r),
\end{equation}
which relates the tunneling functions to the derivative structure that will appear in the squared Hamiltonian. The diagonal block $M$ contains the same-sublattice interlayer tunneling,
\begin{equation}
M(\bm r)=\kappa\begin{pmatrix}0 & \alpha U_0(\bm r)\\ \alpha U_0(-\bm r) & 0\end{pmatrix},
\label{eq:M}
\end{equation}
where $\kappa=w_{AA}/w_{AB}$. The parameter $\kappa$ measures the deviation from the chiral limit. When $\kappa=0$, the same-sublattice tunneling vanishes and the Hamiltonian anticommutes with the chiral operator. For finite $\kappa$, the operator $M$ is diagonal in the sublattice-block structure of Eq.~\eqref{eq:H} but off diagonal in layer space. It therefore represents tunneling between identical sublattices in opposite layers, namely $A_1\leftrightarrow A_2$ and $B_1\leftrightarrow B_2$. In realistic relaxed structures, AA tunneling is reduced but not eliminated, and values close to $\kappa\approx0.7$ are commonly used to describe relaxation-renormalized TBG~\cite{Ledwith2021}.

The moir\'e coupling vectors entering the interlayer potentials are
\begin{equation}
\begin{aligned}
\bm q_1&=k_\theta(0,-1),\\
\bm q_2&=k_\theta\left(\frac{\sqrt3}{2},\frac12\right),\\
\bm q_3&=k_\theta\left(-\frac{\sqrt3}{2},\frac12\right),
\end{aligned}
\end{equation}
where $k_\theta=2k_D\sin(\theta/2)$ is the characteristic moir\'e momentum scale and $k_D=4\pi/(3a_0)$ is the magnitude of the graphene Dirac momentum, with $a_0$ the monolayer lattice constant. Since $k_\theta$ decreases with decreasing twist angle, the moir\'e length scale grows rapidly at small angles.

The continuum model is controlled by the dimensionless coupling parameter
\begin{equation}
\alpha=\frac{w_1}{v_0k_\theta},
\end{equation}
where $w_1\equiv w_{AB}$ and $v_0$ is the Fermi velocity of monolayer graphene. Throughout this work we use units such that $v_0=1$ and $k_\theta=1$, so that lengths are measured in units of $k_\theta^{-1}$ and energies in units of $v_0k_\theta$. In these units, the twist-angle dependence enters through $\alpha$: decreasing the twist angle increases $\alpha$ and strengthens the effective interlayer hybridization relative to the intralayer kinetic energy. In reciprocal space, the moir\'e Brillouin zone is generated by the reciprocal lattice vectors
\begin{equation}
\begin{aligned}
\bm b_{1,2}&=\bm q_{2,3}-\bm q_1=\left(\pm\frac{\sqrt3}{2},\frac32\right),\\
\bm b_3&=\bm q_2-\bm q_3=\big(\sqrt3,0\big).
\end{aligned}
\label{eq:b}
\end{equation}
The high-symmetry points used below are
\begin{equation}
\Gamma=\bm q_1,\qquad K=(0,0),\qquad K'=-\bm q_1,\qquad M=-\frac{\bm q_1}{2},
\end{equation}
following the conventions of Ref.~\cite{Navarro2022}. The corresponding moir\'e real-space lattice vectors can be written compactly as
\begin{equation}
\bm a_{1,2}=\frac{4\pi}{3k_\theta}\left(\pm\frac{\sqrt3}{2},\frac12\right),
\label{eq:a}
\end{equation}
with the sign chosen consistently with $\bm b_{1,2}$. In this gauge, the tunneling functions acquire phase factors under primitive moir\'e translations, while translations by three primitive periods leave them invariant:
\begin{equation}
U_m(\bm r+3\bm a_{1,2})=U_m(\bm r).
\end{equation}
This enlarged periodicity is closely related to the magnetic-translation structure encountered in the quantum-Hall interpretation of the chiral model~\cite{Navarro2023RMF}. It also provides a useful way to organize the Fourier harmonics entering the numerical diagonalization.

The BM Hamiltonian possesses crystalline and antiunitary symmetries that strongly constrain its spectrum and eigenstates. As discussed in Refs.~\cite{Tarnopolsky2019b,Wang2021,Popov2021,Popov2023b,Ceferino2024}, the single-valley continuum model is constrained by $C_{3z}$, $C_{2x}$, $C_{2y}T$, and $C_{2z}T$, where $T$ denotes time reversal. These symmetries restrict the band degeneracies, the valley structure, and the topology of the low-energy bands. The interlayer potentials satisfy
\begin{equation}
U_0^*(\bm r)=U_0(-\bm r),
\end{equation}
while
\begin{equation}
U_1^*(\bm r)=U_{-1}(-\bm r)\neq U_1(-\bm r).
\end{equation}
As a consequence,
\begin{equation}
D^\dagger(\bm r)=D^*(-\bm r),\qquad M^*(-\bm r)=M(\bm r).
\end{equation}
Within the symmetry convention adopted here, the eigenstates can therefore be chosen so that the two sublattice sectors are related by inversion and complex conjugation,
\begin{equation}
\psi_j(\bm r)=\chi_j^*(-\bm r).
\label{eq:symrel}
\end{equation}
This relation will be used later when comparing expectation values computed in the $\Psi$ and $\chi$ sectors.

In the exact chiral limit, the energy spectrum is symmetric with respect to $E=0$. If $|\Phi_+\rangle=(\psi_1,\psi_2,\chi_1,\chi_2)^T$ is an eigenstate with energy $E$, then $|\Phi_-\rangle=(\psi_1,\psi_2,-\chi_1,-\chi_2)^T$ is an eigenstate with energy $-E$. The zero-energy states are therefore special. They satisfy
\begin{equation}
D^\dagger\Psi=0,\qquad D\chi=0,
\end{equation}
and, because they are degenerate, they can be chosen to be polarized in one sublattice sector:
\begin{equation}
|\Phi\rangle=\big(\psi_1(\bm r),\psi_2(\bm r),0,0\big)^T,
\end{equation}
or alternatively,
\begin{equation}
|\Phi\rangle=\big(0,0,\chi_1(\bm r),\chi_2(\bm r)\big)^T.
\end{equation}
At the chiral magic angles, these zero modes form the completely flat bands of the TKV model. Their existence is protected by the chiral structure of the Hamiltonian and by the topology of the zero-mode problem. They have also been connected to topological solitons protected by crystalline symmetries and to compact localized states in flat-band lattice systems~\cite{Andrade2023,DeBeule2021,Naumis1994}. The finite-$\kappa$ problem studied in the following sections can therefore be viewed as a deformation of this chiral zero-mode structure. Same-sublattice tunneling breaks exact flatness, but part of the symmetry and topology inherited from the chiral limit remains visible in the low-energy bands.

\section{Spectral squeezing and wave-function localization}

\begin{figure*}[t]
\centering
\includegraphics[width=\textwidth]{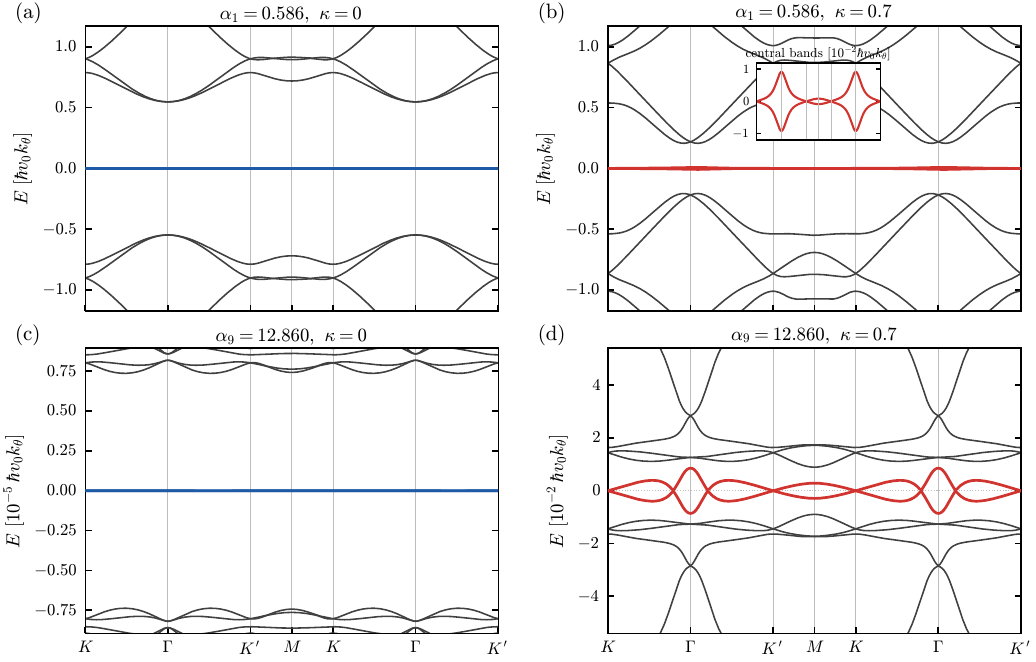}
\caption{Band structure for the first magic angle, panels (a) and (b), and for a higher-order magic angle, panels (c) and (d), comparing the chiral model with the relaxation-renormalized non-chiral model at $\kappa=0.7$. In the chiral model, the bands squeeze as a function of $\alpha$, and the gap decreases exponentially as predicted by the chiral scaling theory. In the non-chiral model, the hybridization between the central bands and the first remote bands becomes substantially stronger, especially at small twist angles where the gap tends to close systematically.}
\label{fig:bands}
\end{figure*}

We now analyze the spectrum and wave functions obtained from the continuum Hamiltonian in Eq.~\eqref{eq:H}. Figure~\ref{fig:bands} compares the chiral limit, $\kappa=0$, with the relaxation-renormalized non-chiral model, $\kappa=0.7$, for representative values of the coupling parameter $\alpha$ corresponding to the first and higher-order magic-angle regimes. This comparison isolates the effect of finite same-sublattice tunneling on the low-energy bands. In the perturbative regime $\kappa\alpha\ll1$, the numerical results can also be compared with the analytical expansion discussed in the Appendix.

\begin{figure}[t]
\centering
\includegraphics[width=\columnwidth]{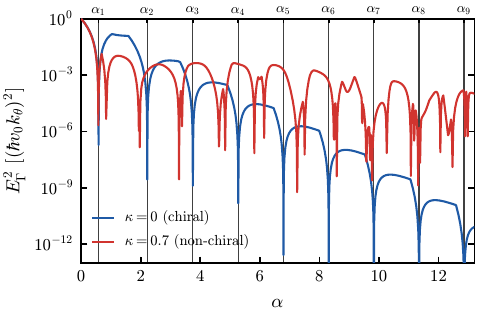}
\caption{$E^2(\alpha)$ at the $\Gamma$ point in semi-logarithmic scale for the non-chiral model $\kappa\approx0.7$ (red curve), compared with the chiral model $\kappa=0$ (blue curve). The vertical black lines indicate the magic angles of the chiral theory. While the chiral model preserves an exponential squeezing of the spectrum, finite $\kappa$ partially blocks this behavior and produces substantial deviations at higher-order magic angles.}
\label{fig:E2}
\end{figure}

A first diagnostic is provided by the behavior of the band-edge energy at the $\Gamma$ point. In Fig.~\ref{fig:E2} we plot $E^2$ as a function of $\alpha$ for the low-energy state that controls the band edge in the numerical spectrum. Together with the band structures in Fig.~\ref{fig:bands}, this quantity shows the qualitative difference between the chiral and non-chiral regimes. In the exact chiral limit, the bandwidths $W_n$ associated with the low-energy bands follow the characteristic squeezing behavior
\begin{equation}
W_n\sim\alpha^{-1}.
\end{equation}
This progressive narrowing of the spectrum is one of the central signatures of the chiral model. The origin of this squeezing mechanism can be understood from the real-space rescaling introduced in the chiral theory~\cite{Navarro2023},
\begin{equation}
\bm r\to\bm r'=\frac{\bm r}{\alpha}.
\end{equation}
This transformation exposes the quantum-Hall-like structure of the squared chiral Hamiltonian in the large-$\alpha$ regime~\cite{Navarro2023,Navarro2024}. In this limit, corrections associated with the effective Zeeman contribution become asymptotically small~\cite{Navarro2023,Navarro2022,Navarro2024}. Under the same scaling, the dominant operators in the non-chiral Hamiltonian behave as
\begin{equation}
D\sim\alpha,\qquad M\sim\alpha\kappa.
\end{equation}
Therefore, finite $\kappa$ introduces terms that scale with the same power of $\alpha$ as the chiral interlayer coupling. These terms compete directly with the chiral scaling structure and obstruct the exact renormalization mechanism responsible for ideal spectral squeezing. This obstruction is visible already in the low-energy bands. In the chiral model, exact flatness is recovered at the magic angles of the TKV theory, while away from those values the bands display the expected squeezing with increasing $\alpha$. Once finite $\kappa$ is included, the central bands acquire residual dispersion and the higher-order magic-angle structure is strongly modified. The magic-angle values are also shifted relative to the chiral prediction. Thus, the realistic model does not preserve the exact flatness of the ideal chiral continuum theory, in agreement with symmetry-based analyses showing that atomic-scale models are not expected to retain the full idealized flat-band structure~\cite{Song2022}. The effect is especially pronounced near the $\Gamma$ point, where the band edge is located in the spectra considered here. For the first magic angle, the non-chiral bands remain narrow and close to the chiral result. At higher-order magic angles, however, finite $\kappa$ produces a much stronger deviation: the squeezing of the low-energy spectrum is partially blocked, and the separation between the central bands and the first remote bands is strongly reduced. This indicates that remote-band effects become increasingly important at small twist angles and that isolated-flat-band descriptions become less reliable in this regime.

The low-energy states near the high-symmetry points $\bm K$ and $\bm K'$ remain close to charge neutrality even when $\kappa$ is finite. This behavior reflects the fact that the same-sublattice tunneling term preserves the crystalline symmetries of the continuum model while breaking the exact chiral anticommutation symmetry. Hence, part of the symmetry and topology inherited from the chiral limit remains visible in the non-chiral spectrum, although the exact spectral symmetry around zero energy is lost.

\begin{figure*}[t]
\centering
\includegraphics[width=0.66\textwidth]{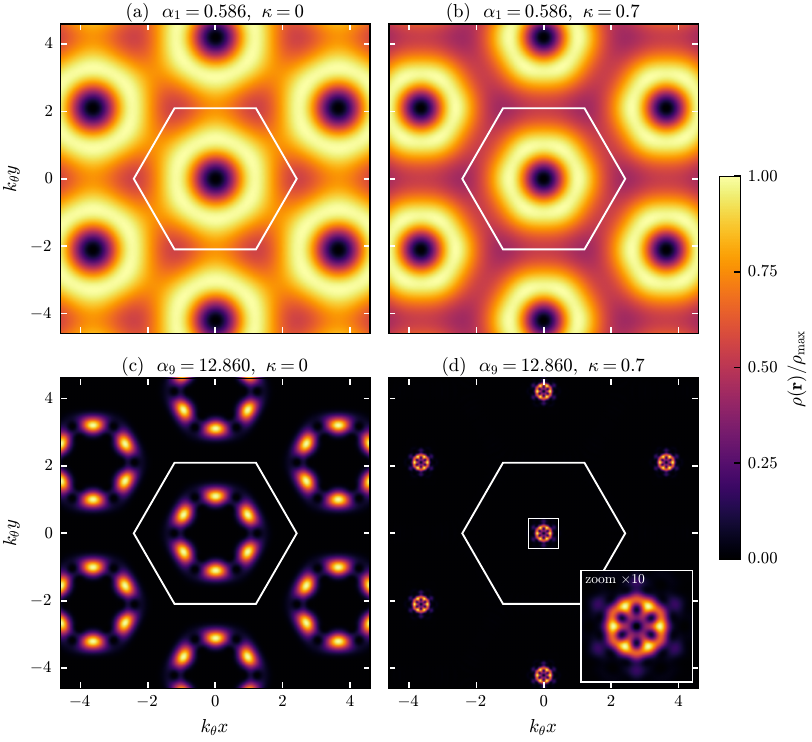}
\caption{Electronic density at the $\bm\Gamma$ point. First magic angle $\alpha_1=0.586$ for (a) the chiral limit $\kappa=0$ and (b) the non-chiral model with $\kappa=0.7$. Electronic density for the higher-order magic angle $\alpha_9=12.860$ for (c) the chiral limit $\kappa=0$ and (d) the non-chiral model $\kappa=0.7$. Finite $\kappa$ induces a pronounced localization of the electronic density around the AA stacking regions.}
\label{fig:density}
\end{figure*}

A second consequence of finite $\kappa$ is the enhanced hybridization between the central bands and the first remote bands. In the chiral model, the central bands are protected by the zero-mode structure and remain well separated from the remote bands at the magic angles. In the non-chiral model, AA tunneling opens additional same-sublattice tunneling channels and mixes states that are effectively decoupled in the chiral limit. As shown in Fig.~\ref{fig:bands}, this mixing becomes particularly strong at larger $\alpha$, where the remote-band gap becomes very small and the central bands are no longer cleanly isolated from the first excited bands. The corresponding real-space effect is shown in Fig.~\ref{fig:density}. At the first magic angle, the chiral model displays a broad and symmetric density distribution around the AA regions. When finite $\kappa$ is introduced, the density profile becomes narrower, indicating stronger localization. The underlying $C_3$ symmetry is preserved, but the electronic weight contracts toward the center of the AA stacking region. This localization becomes much more pronounced at higher-order magic angles. For $\alpha_9=12.860$, the non-chiral density is concentrated in a substantially smaller spatial region than in the chiral case. Finite same-sublattice tunneling therefore acts as an effective localization mechanism for the low-energy wave functions. This enhanced localization is expected to increase local interaction matrix elements and may modify the balance between kinetic-energy suppression and Coulomb interactions in realistic moir\'e systems.

\begin{figure*}[t]
\centering
\includegraphics[width=0.66\textwidth]{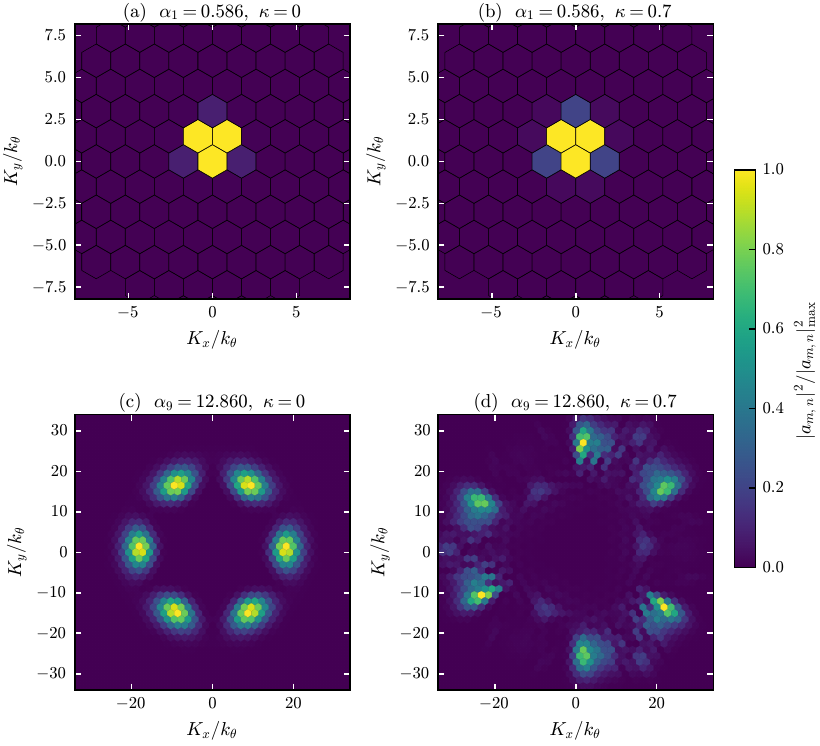}
\caption{Fourier coefficients $|a_{m,n}|^2$ for the chiral (left column) and non-chiral (right column) models. Results are shown for the first magic angle, $\alpha_1=0.586$, and for the higher-order magic angle, $\alpha_9=12.860$. The non-chiral calculations use $\kappa=0.7$, following previous estimates for relaxation-renormalized TBG~\cite{Ledwith2022}. Finite $\kappa$ shifts spectral weight toward larger $|\bm K_{mn}|$, indicating stronger localization in real space. It also enhances the trigonal anisotropy of the Fourier distribution while preserving the underlying $C_3$ symmetry.}
\label{fig:fourier}
\end{figure*}

The real-space localization is consistent with the redistribution of Fourier weight shown in Fig.~\ref{fig:fourier}. We plot the coefficients $|a_{mn}|^2$ for the chiral and non-chiral models at $\alpha_1=0.586$ and $\alpha_9=12.860$. At the first magic angle, the two distributions remain qualitatively similar: most of the weight is concentrated around a small number of symmetry-related reciprocal-space points. Finite $\kappa$ nevertheless changes the relative intensity and angular distribution of the dominant components. At the higher-order magic angle, the difference becomes much stronger. The dominant Fourier components move to larger $|\bm K_{mn}|$ in the non-chiral model, indicating that the real-space wave function is more tightly localized. The approximately sixfold pattern observed in the chiral limit evolves into a distribution with stronger trigonal anisotropy, while the underlying $C_3$ symmetry of the continuum Hamiltonian remains preserved. Thus, finite $\kappa$ does not simply broaden the bands; it also reorganizes the wave functions in reciprocal space and concentrates the electronic density in real space.

Overall, finite same-sublattice tunneling qualitatively modifies the chiral flat-band hierarchy. It suppresses the ideal spectral squeezing, enhances hybridization between the central and remote bands, and localizes the low-energy wave functions more strongly around AA regions. These effects show that the realistic BM model remains continuously connected to the chiral theory, but with important finite-$\kappa$ corrections that become increasingly relevant at higher-order magic angles. In the next section, we analyze these corrections directly from the squared Hamiltonian.

\section{Squared TBG Hamiltonian with finite-$\kappa$ corrections}

The previous section showed that finite same-sublattice tunneling suppresses the chiral spectral squeezing, enhances hybridization with remote bands, and localizes the low-energy wave functions more strongly around the AA regions. We now analyze the origin of these effects from the squared Hamiltonian. In the chiral limit, the squared formulation reduces the problem to an effective $2\times2$ operator, where the relation between flat bands, confinement, and emergent pseudo-magnetic fields becomes transparent~\cite{Navarro2023,Navarro2022,Navarro2023RMF,Naumis2021,Navarro2024}. Starting from Eq.~\eqref{eq:H}, the square of the non-chiral Hamiltonian is
\begin{equation}
\mathcal H^2=\begin{pmatrix} D^\dagger D+M^2 & \{M,D^\dagger\}\\ \{M,D\} & DD^\dagger+M^2\end{pmatrix},
\label{eq:H2}
\end{equation}
where $\{M,D\}$ denotes the anticommutator between $M$ and $D$. In the exact chiral limit, $M=0$, the anticommutator terms vanish and the two sublattice sectors decouple. The squared Hamiltonian then consists of the partner operators $D^\dagger D$ and $DD^\dagger$, whose zero modes generate the flat bands of the chiral model.

For finite $\kappa$, this decoupling is no longer exact. Equation~\eqref{eq:H2} contains two distinct non-chiral corrections. The term $M^2$ modifies the diagonal confinement potential, while the anticommutators $\{M,D\}$ and $\{M,D^\dagger\}$ couple the two sectors of the squared problem. These two contributions provide the microscopic mechanism behind the numerical trends observed in Fig.~\ref{fig:bands}: the loss of ideal squeezing, the enhanced central--remote-band hybridization, and the stronger real-space localization.

The Schr\"odinger equations associated with Eq.~\eqref{eq:H2} are
\begin{equation}
(D^\dagger D+M^2)\Psi+\{M,D^\dagger\}\chi=E^2\Psi,
\label{eq:Sch1}
\end{equation}
and
\begin{equation}
(DD^\dagger+M^2)\chi+\{M,D\}\Psi=E^2\chi.
\label{eq:Sch2}
\end{equation}
Equation~\eqref{eq:Sch2} can be formally solved for $\chi$ and substituted into Eq.~\eqref{eq:Sch1}. This gives an effective equation for the $\Psi$ sector,
\begin{multline}
(D^\dagger D+M^2)\Psi\\
-\{M,D^\dagger\}(DD^\dagger+M^2-E^2)^{-1}\{M,D\}\Psi=E^2\Psi.
\label{eq:Seff}
\end{multline}
This expression shows that the non-chiral model cannot be reduced to the chiral squared Hamiltonian by a simple parameter renormalization. The diagonal term $M^2$ changes the effective confinement, while the second term in Eq.~\eqref{eq:Seff} produces an energy-dependent coupling to the opposite sublattice sector. A perturbative expansion is possible in the regime $\kappa\alpha\ll1$, as discussed in the Appendix, and provides a useful check of the numerical results at small $\alpha$. For the realistic value $\kappa\simeq0.7$, however, the full squared-Hamiltonian structure must be retained.

It is useful to separate the diagonal and off-diagonal parts of Eq.~\eqref{eq:H2} as
\begin{equation}
\mathcal H^2=\mathcal H_f^2+\begin{pmatrix}0 & \{M,D^\dagger\}\\ \{M,D\} & 0\end{pmatrix},
\label{eq:H2split}
\end{equation}
where
\begin{equation}
\mathcal H_f^2=\begin{pmatrix}H_f^2 & 0\\ 0 & \widetilde H_f^2\end{pmatrix}.
\end{equation}
The two diagonal blocks are
\begin{equation}
H_f^2=H^2+M^2,\qquad \widetilde H_f^2=\widetilde H^2+M^2,
\end{equation}
with
\begin{equation}
H^2=D^\dagger D,\qquad \widetilde H^2=DD^\dagger.
\end{equation}
Thus $H^2$ and $\widetilde H^2$ are the two squared chiral partner Hamiltonians, while $M^2$ gives the diagonal finite-$\kappa$ correction. The squaring procedure is closely related to supersymmetric mappings of bipartite Hamiltonians~\cite{Matsumoto2023,Mizoguchi2020,Mizoguchi2023,Yoshida2021}; in the present context, its main advantage is that it exposes the confinement and pseudo-magnetic structure of the chiral problem~\cite{Navarro2023,Navarro2022}.

To keep the matrix expressions compact, we define the diagonal operators
\begin{equation}
\mathcal V_\pm[X]=-\nabla^2+\alpha^2\left[X(\bm r)+\Delta(\pm\bm r)\right],
\label{eq:Vpm}
\end{equation}
where $X(\bm r)$ denotes the symmetric part of the diagonal confinement. We also define
\begin{equation}
\mathcal J_+=\alpha(-2i\bm A\cdot\nabla+B),\qquad \mathcal J_-=\alpha(-2i\bm A^*\cdot\nabla+B^*).
\label{eq:Jpm}
\end{equation}
The notation $\mathcal J_-$ denotes the lower-left operator appearing in the squared Hamiltonian; it should not be interpreted as the complex conjugate of $\mathcal J_+$, since the factor $-2i$ is kept fixed by the operator structure. Hermiticity is understood with the derivative operators acting on periodic functions in the moir\'e unit cell; $\mathcal J_-$ is the lower-left differential operator in this convention, not the naive complex conjugate of $\mathcal J_+$.

With this notation, the squared chiral Hamiltonian takes the form~\cite{Navarro2024b}
\begin{equation}
H^2=\begin{pmatrix}\mathcal V_+[\bm A^2] & \mathcal J_+\\ \mathcal J_- & \mathcal V_-[\bm A^2]\end{pmatrix}.
\end{equation}
The complex vector quantity entering Eq.~\eqref{eq:Jpm} is
\begin{equation}
\bm A=\sum_{\nu=1}^{3}e^{-i\bm q_\nu\cdot\bm r}\bm q_\nu^\perp.
\end{equation}
It acts as an emergent pseudo-magnetic vector potential. It satisfies the Coulomb gauge condition $\nabla\cdot\bm A=0$ and generates an effective field perpendicular to the graphene plane,
\begin{equation}
\bm B=\nabla\times\bm A,\qquad \bm B^*=\nabla\times\bm A^*.
\end{equation}
We denote the corresponding out-of-plane scalar component by $B$, which in the present convention is
\begin{equation}
B=-i\sum_\nu e^{-i\bm q_\nu\cdot\bm r}=-iU_0(\bm r).
\end{equation}
The complex structure of this field follows from the valley structure and the phase winding generated by the moir\'e tunneling functions~\cite{Navarro2023RMF}.

The diagonal potentials of the chiral squared Hamiltonian can be decomposed into a symmetric confinement term and an antisymmetric pseudo-magnetic contribution,
\begin{equation}
\begin{aligned}
\bm A^2(\bm r)&=3-\sum_{\nu=1}^{3}\cos(\bm b_\nu\cdot\bm r),\\
\Delta(\bm r)&=i[A_x,A_y]=\sqrt3\sum_{\nu=1}^{3}(-1)^\nu\sin(\bm b_\nu\cdot\bm r).
\end{aligned}
\label{eq:A2Delta}
\end{equation}
Thus $\bm A^2(\bm r)$ defines the symmetric confinement landscape, while $\Delta(\bm r)$ encodes the antisymmetric part of the effective pseudo-magnetic structure. We now include the diagonal non-chiral correction $M^2$. Since
\begin{equation}
H_f^2=H^2+M^2,
\end{equation}
finite $\kappa$ renormalizes the symmetric confinement term according to
\begin{equation}
\bm A^2\to\bm A_f^2,
\end{equation}
with
\begin{equation}
\bm A_f^2=3(1+\kappa^2)+(-1+2\kappa^2)\sum_\mu\cos(\bm b_\mu\cdot\bm r).
\label{eq:Af2}
\end{equation}
This expression explains why finite $\kappa$ changes the localization properties found in the previous section: the same-sublattice tunneling does not only break chiral symmetry, but also reshapes the effective confinement potential experienced by the low-energy states. Using Eq.~\eqref{eq:Vpm}, the finite-$\kappa$ block becomes
\begin{equation}
H_f^2=\begin{pmatrix}\mathcal V_+[\bm A_f^2] & \mathcal J_+\\ \mathcal J_- & \mathcal V_-[\bm A_f^2]\end{pmatrix}.
\label{eq:Hf2}
\end{equation}
An especially important reference point occurs for
\begin{equation}
\kappa=\frac{1}{\sqrt2},
\end{equation}
which is strikingly close to the commonly used relaxation-renormalized value $\kappa\simeq0.7$. At this value, the oscillatory cosine contribution in Eq.~\eqref{eq:Af2} vanishes exactly. Therefore, finite same-sublattice tunneling does not merely perturb the chiral confinement landscape; near the realistic value of $\kappa$, it cancels the spatial modulation of the symmetric confinement term and leaves a uniform contribution,
\begin{equation}
\alpha^2\bm A_f^2=\frac{(3\alpha)^2}{2}=\frac{1}{2\sigma^4},
\end{equation}
where
\begin{equation}
\sigma=\frac{1}{\sqrt{3\alpha}}.
\end{equation}
Here $\sigma$ is the localization scale obtained from the asymptotic scaling analysis~\cite{Navarro2023}. This cancellation provides a direct analytical interpretation of the relaxation-renormalized BM parameter: the realistic non-chiral model lies close to a special confinement point of the squared Hamiltonian. The fact that realistic TBG samples appear naturally close to this special value suggests the intriguing possibility that lattice relaxation favors a nearly optimized confinement regime, in which the oscillatory part of the symmetric potential is strongly suppressed. The remaining spatial structure of the diagonal block is then controlled by the antisymmetric pseudo-magnetic term $\Delta(\bm r)$, while the off-diagonal anticommutator terms discussed below provide the additional coupling between the two squared-Hamiltonian sectors. Under the approximation $\kappa\approx1/\sqrt2$, the diagonal block simplifies to
\begin{equation}
H_f^2\approx\begin{pmatrix}\mathcal V_+[9/2] & \mathcal J_+\\ \mathcal J_- & \mathcal V_-[9/2]\end{pmatrix}.
\end{equation}

We finally analyze the off-diagonal blocks in Eq.~\eqref{eq:H2split}. These terms are absent in the chiral limit and are responsible for coupling the two squared-Hamiltonian sectors. Direct evaluation gives
\begin{equation}
\{M,D\}=\kappa\alpha^2V_S(\bm r)\tau_0-i\kappa\alpha\begin{pmatrix}0 & \{B^*,p\}\\ -\{B,p\} & 0\end{pmatrix},
\label{eq:MD}
\end{equation}
where $p=p_x-ip_y$, with $p_j=-i\partial_j$ in the dimensionless units used here. Similarly,
\begin{equation}
\{M,D^\dagger\}=\kappa\alpha^2V_S^\dagger(\bm r)\tau_0-i\kappa\alpha\begin{pmatrix}0 & \{B^*,p^*\}\\ -\{B,p^*\} & 0\end{pmatrix}.
\label{eq:MDd}
\end{equation}
The scalar potential
\begin{equation}
V_S(\bm r)=U_0(\bm r)U_1(-\bm r)+U_0(-\bm r)U_1(\bm r)
\end{equation}
is a complex sublattice-sector coupling generated by finite same-sublattice tunneling. Explicitly,
\begin{equation}
\begin{aligned}
V_S(\bm r)={}&2(1+e^{i\phi})\cos(\bm b_1\cdot\bm r)\\
&+2(1+e^{-i\phi})\cos(\bm b_2\cdot\bm r)\\
&-2\cos(\bm b_3\cdot\bm r).
\end{aligned}
\label{eq:VS}
\end{equation}
The first term in Eqs.~\eqref{eq:MD} and \eqref{eq:MDd} scales as $\kappa\alpha^2$, while the momentum-dependent term scales as $\kappa\alpha$. These off-diagonal contributions therefore become increasingly relevant at larger $\alpha$, precisely where the numerical spectra show stronger remote-band hybridization and a breakdown of the ideal chiral squeezing.

\begin{figure*}[t]
\centering
\includegraphics[width=\textwidth]{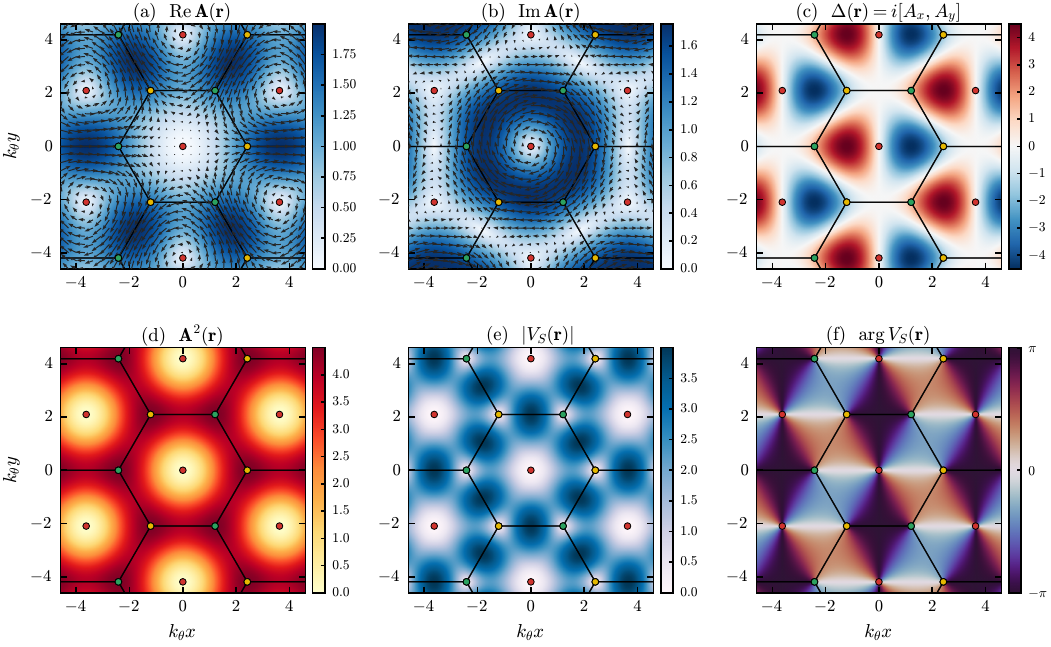}
\caption{Effective potentials of the non-chiral squared Hamiltonian in real space. Panels (a) and (b) show the real and imaginary components of the intrinsic pseudo-magnetic vector potential $\bm A(\bm r)$, respectively. Panels (c) and (d) display the antisymmetric confinement potential $\Delta(\bm r)=i[A_x,A_y]$ and the symmetric confinement term $\bm A^2(\bm r)$. The non-chiral nature of the model introduces the complex sublattice-sector coupling potential $V_S(\bm r)$, whose modulus and phase are shown in panels (e) and (f), respectively. The moir\'e unit cell is indicated together with the AA stacking regions (red points) and the AB/BA stacking regions (yellow and green points).}
\label{fig:potentials}
\end{figure*}

Figure~\ref{fig:potentials} summarizes the real-space structure of the effective fields entering the squared Hamiltonian. The symmetric potential controls the confinement landscape, the antisymmetric term carries the pseudo-magnetic structure, and the complex potential $V_S(\bm r)$ represents the additional sublattice-sector coupling generated by finite $\kappa$. Thus the two main effects found numerically have a direct origin in the squared Hamiltonian: $M^2$ renormalizes the confinement landscape and changes the localization scale, while the anticommutator terms couple the chiral sectors and enhance hybridization with remote bands.

\begin{figure*}[t]
\centering
\includegraphics[width=\textwidth]{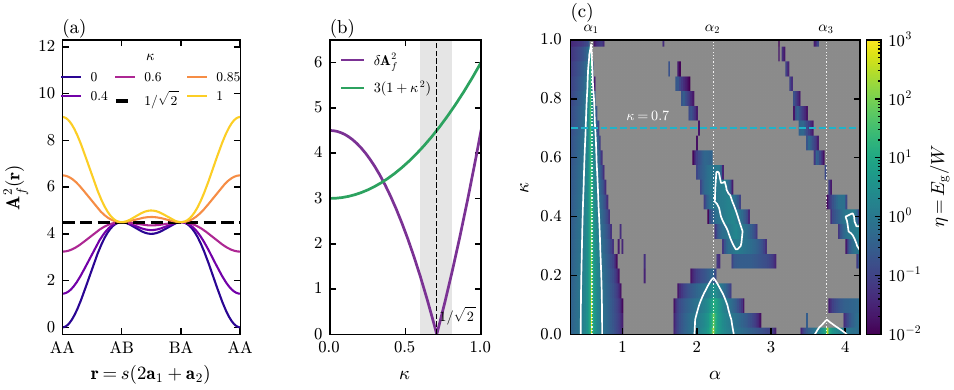}
\caption{Confinement uniformization and spectral isolation of the magic angles. (a) Symmetric confinement $\bm A_f^2$ along the path $\bm r=s(2\bm a_1+\bm a_2)$ connecting two AA regions through the AB and BA domains, Eq.~\eqref{eq:Afcut}, for several values of $\kappa$. All curves cross at the AB/BA value $9/2$; at $\kappa=1/\sqrt2$ (dashed line) the profile is flat. (b) Peak-to-peak modulation $\delta\bm A_f^2$ of Eq.~\eqref{eq:dAf} and uniform part $3(1+\kappa^2)$ as functions of $\kappa$. The shaded band marks the relaxation-renormalized interval $0.6\lesssim\kappa\lesssim0.8$. (c) Isolation ratio $\eta=E_{\rm g}/W$ of the two central bands, Eq.~\eqref{eq:Wgap}, in the $(\alpha,\kappa)$ plane. Gray regions correspond to $\eta\le10^{-2}$, including the cases in which the central bands overlap in energy with the remote bands ($E_{\rm g}\le0$). The white contour marks $\eta=1$, the dotted lines indicate the chiral magic angles $\alpha_{1,2,3}$, and the dashed line indicates $\kappa=0.7$.}
\label{fig:confinement}
\end{figure*}

The confinement mechanism and its spectral consequences can be made explicit with the diagnostics summarized in Fig.~\ref{fig:confinement}. Along the straight path that joins two neighboring AA regions through the AB and BA domains, $\bm r(s)=s(2\bm a_1+\bm a_2)$ with $0\le s\le1$, the vectors in Eqs.~\eqref{eq:b} and \eqref{eq:a} give $\bm b_1\cdot\bm r=4\pi s$ and $\bm b_2\cdot\bm r=\bm b_3\cdot\bm r=2\pi s$, so that Eq.~\eqref{eq:Af2} reduces to the one-dimensional profile
\begin{equation}
\bm A_f^2(s)=3(1+\kappa^2)+(2\kappa^2-1)\left[\cos(4\pi s)+2\cos(2\pi s)\right].
\label{eq:Afcut}
\end{equation}
At the AB and BA points, $s=1/3$ and $s=2/3$, the bracket equals $-3/2$ and $\bm A_f^2=9/2$ independently of $\kappa$, whereas at the AA points, $s=0$ and $s=1$, one finds $\bm A_f^2=9\kappa^2$. Finite $\kappa$ therefore lifts the AA minimum of the chiral confinement, $\bm A^2(\bm 0)=0$, until it reaches the AB/BA level exactly at $\kappa=1/\sqrt2$; for $\kappa>1/\sqrt2$ the AA regions turn into maxima of the symmetric confinement [Fig.~\ref{fig:confinement}(a)]. Since $-3/2\le\sum_\mu\cos(\bm b_\mu\cdot\bm r)\le3$ over the whole moir\'e cell, the peak-to-peak modulation of the symmetric confinement is
\begin{equation}
\delta\bm A_f^2\equiv\max_{\bm r}\bm A_f^2-\min_{\bm r}\bm A_f^2=\frac92\left|2\kappa^2-1\right|,
\label{eq:dAf}
\end{equation}
which vanishes linearly at $\kappa=1/\sqrt2$, while the uniform part $3(1+\kappa^2)$ grows smoothly [Fig.~\ref{fig:confinement}(b)]. For $\kappa=0.7$, $\delta\bm A_f^2$ is only $2\%$ of its chiral value $\delta\bm A^2=9/2$.

To connect this confinement landscape with the spectral robustness of the magic angles, we characterize the two central bands $E_{\pm1}(\bm k)$ by their total bandwidth and by their separation from the first remote bands $E_{\pm2}(\bm k)$,
\begin{equation}
\begin{aligned}
W&=\max_{\bm k}E_{+1}(\bm k)-\min_{\bm k}E_{-1}(\bm k),\\
E_{\rm g}&=\min\Big[\min_{\bm k}E_{+2}-\max_{\bm k}E_{+1},\;\min_{\bm k}E_{-1}-\max_{\bm k}E_{-2}\Big],
\end{aligned}
\label{eq:Wgap}
\end{equation}
and define the isolation ratio $\eta=E_{\rm g}/W$. An isolated flat band requires $\eta\gg1$, whereas $E_{\rm g}\le0$ signals that the central bands overlap in energy with remote bands. Figure~\ref{fig:confinement}(c) shows $\eta$ in the $(\alpha,\kappa)$ plane, with the extrema in Eq.~\eqref{eq:Wgap} evaluated along the closed path $\Gamma$--$K$--$M$--$K'$--$\Gamma$ of the moir\'e Brillouin zone. In the chiral limit, the first three magic angles appear as narrow regions with $\eta\gg1$. The region around $\alpha_1$ remains isolated, $\eta>1$, up to $\kappa\approx0.9$, and for $\kappa=0.7$ the isolation ratio still reaches $\eta\simeq26$ at $\alpha\simeq0.575$, slightly below the chiral value. By contrast, the regions associated with $\alpha_2$ and $\alpha_3$ close already for $\kappa\lesssim0.2$; beyond that point only narrow, shifted pockets survive at intermediate $\kappa$, and for $\kappa=0.7$ one finds $\eta<0.5$ in the vicinity of both $\alpha_2$ and $\alpha_3$. Figure~\ref{fig:confinement}(c) therefore provides a compact spectral summary of the robustness of the first magic angle and of the destabilization of the higher-order members of the chiral hierarchy, consistent with Figs.~\ref{fig:bands} and \ref{fig:E2}.

\section{Energetic contributions in the non-chiral model}

In this section, we analyze the energetic contributions associated with the different terms entering the non-chiral squared Hamiltonian. The expectation values provide direct information about the physical mechanisms governing the localization of the electronic states, the suppression of the bandwidth, and the redistribution of interlayer processes generated by finite same-sublattice tunneling. Starting from Eq.~\eqref{eq:H2}, the expectation value of the squared Hamiltonian is
\begin{multline}
\langle\Psi|(D^\dagger D+M^2)|\Psi\rangle+\langle\Psi|\{M,D^\dagger\}|\chi\rangle\\
+\langle\chi|(DD^\dagger+M^2)|\chi\rangle+\langle\chi|\{M,D\}|\Psi\rangle=E^2.
\label{eq:expH2}
\end{multline}
The two diagonal sectors are related by the symmetry relation in Eq.~\eqref{eq:symrel} for the states considered here. Therefore, the expectation values can be evaluated using one sector and then compared with the corresponding symmetry-related contribution from the other sector. In the numerical results below, all plotted quantities are reported using the same normalization convention, so that the relative weights of the kinetic, confinement, pseudo-magnetic, and interlayer terms can be directly compared.

Using the explicit operators introduced in the previous section, the energetic decomposition can be written as
\begin{multline}
\langle T\rangle+\alpha^2\left(\langle\bm A_f^2\rangle+\langle\Delta\rangle\right)\\
-\alpha\left(\langle\bm A\cdot\nabla\rangle+\langle\bm B\cdot\bm\tau\rangle+\kappa\langle A_{g,s}\rangle\right)+\kappa\alpha^2\langle V_S\rangle=E^2.
\label{eq:decomp}
\end{multline}
Here $\Delta(\bm r)=i[A_x,A_y]$, and the signs are inherited from the squared-Hamiltonian convention used in Eq.~\eqref{eq:Hf2} and in the anticommutator terms of Eqs.~\eqref{eq:MD} and \eqref{eq:MDd}. The first line of Eq.~\eqref{eq:decomp} contains the kinetic and diagonal confinement energies, while the second line contains the chiral orbital interlayer contribution, the pseudo-Zeeman contribution, and the non-chiral inter-sublattice terms activated by finite $\kappa$.

The kinetic contribution is
\begin{equation}
\langle T\rangle=-\langle\Psi|\nabla^2|\Psi\rangle=-\sum_{j=1}^{2}\int_{\rm u.c.}\psi_j^*\nabla^2\psi_j\,d^2\bm r,
\end{equation}
with the symmetry-related contribution from the $\chi$ sector included through the normalization convention described above. The diagonal symmetric confinement term is
\begin{equation}
\langle\bm A_f^2\rangle=\int_{\rm u.c.}\sum_{j=1}^{2}\psi_j^*\bm A_f^2(\bm r)\psi_j\,d^2\bm r.
\end{equation}
Using Eq.~\eqref{eq:Af2}, this becomes
\begin{multline}
\langle\bm A_f^2\rangle=3(1+\kappa^2)\,\mathcal N_\Psi\\
+(-1+2\kappa^2)\sum_{\mu=1}^{3}\sum_{j=1}^{2}\int_{\rm u.c.}\psi_j^*\cos(\bm b_\mu\cdot\bm r)\psi_j\,d^2\bm r,
\label{eq:Af2exp}
\end{multline}
where
\begin{equation}
\mathcal N_\Psi=\sum_{j=1}^{2}\int_{\rm u.c.}|\psi_j(\bm r)|^2\,d^2\bm r
\end{equation}
is the norm of the sector used in the expectation value. If the sector density is normalized to the moir\'e unit-cell area, then $\mathcal N_\Psi=A_M$, with $A_M=8\pi^2/(3\sqrt3)$. Equation~\eqref{eq:Af2exp} is the energetic counterpart of the confinement result derived in the previous section. The uniform part grows as $3(1+\kappa^2)$, while the oscillatory part is controlled by the coefficient $-1+2\kappa^2$. Therefore, at
\begin{equation}
\kappa=\frac{1}{\sqrt2},
\end{equation}
the spatially oscillating contribution vanishes exactly. Since the relaxation-renormalized value $\kappa\simeq0.7$ lies extremely close to this point, realistic TBG is naturally placed near a regime in which the symmetric confinement is dominated by a uniform contribution and the oscillatory modulation is strongly suppressed. In this regime,
\begin{equation}
\langle\bm A_f^2\rangle\simeq\frac92\mathcal N_\Psi,
\end{equation}
or equivalently $\langle\bm A_f^2\rangle\simeq(9/2)A_M$ when the sector normalization is chosen as $\mathcal N_\Psi=A_M$. This is not a minor correction to the chiral model; it identifies a special confinement point selected by the realistic value of the relaxation-renormalized BM parameter.

The remaining diagonal term is the antisymmetric pseudo-magnetic contribution,
\begin{equation}
\begin{aligned}
\langle\Delta\rangle&=\int_{\rm u.c.}\sum_{j=1}^{2}\psi_j^*\Delta(\bm r)\psi_j\,d^2\bm r\\
&=\sqrt3\sum_{\nu=1}^{3}(-1)^\nu\sum_{j=1}^{2}\int_{\rm u.c.}\psi_j^*\sin(\bm b_\nu\cdot\bm r)\psi_j\,d^2\bm r.
\end{aligned}
\end{equation}
This contribution remains bounded because $\Delta(\bm r)$ is a bounded periodic function. Thus, in Eq.~\eqref{eq:decomp}, the diagonal confinement terms enter with an overall prefactor $\alpha^2$. The uniformization of $\bm A_f^2$ near $\kappa\simeq1/\sqrt2$ therefore has a direct energetic consequence: it changes the leading $\alpha^2$ confinement channel while leaving the remaining spatial dependence of the diagonal block mainly controlled by $\Delta(\bm r)$.

The chiral off-diagonal contribution inside the squared Hamiltonian is
\begin{equation}
\begin{aligned}
\langle\bm A\cdot\nabla\rangle&=2i\int_{\rm u.c.}\left(\psi_2^*\bm A\cdot\nabla\psi_1+\psi_1^*\bm A^*\cdot\nabla\psi_2\right)d^2\bm r\\
&=2i\int_{\rm u.c.}\sum_{\nu=1}^{3}\left(\psi_2^*e^{-i\bm q_\nu\cdot\bm r}\nabla_\nu\psi_1+\psi_1^*e^{i\bm q_\nu\cdot\bm r}\nabla_\nu\psi_2\right)d^2\bm r,
\end{aligned}
\label{eq:Agrad}
\end{equation}
where
\begin{equation}
\nabla_\nu=\bm q_\nu^\perp\cdot\nabla.
\end{equation}
This term contains one spatial derivative and therefore becomes increasingly important as the wave functions localize in real space. In the large-$\alpha$ regime, the characteristic gradients increase with the inverse localization length, so the full contribution $-\alpha\langle\bm A\cdot\nabla\rangle$ competes directly with the diagonal $\alpha^2$ confinement energy.

The pseudo-Zeeman contribution is
\begin{equation}
\begin{aligned}
\langle\bm B\cdot\bm\tau\rangle&=\int_{\rm u.c.}\left(\chi_2^*\bm B\cdot\bm\tau\psi_1+\chi_1^*\bm B^*\cdot\bm\tau\psi_2\right)d^2\bm r\\
&=i\int_{\rm u.c.}\sum_{\nu=1}^{3}\left(-\chi_2^*e^{-i\bm q_\nu\cdot\bm r}\psi_1+\chi_1^*e^{i\bm q_\nu\cdot\bm r}\psi_2\right)d^2\bm r.
\end{aligned}
\label{eq:Btau}
\end{equation}
Unlike Eq.~\eqref{eq:Agrad}, this term contains no derivative. It is therefore most visible at the first magic angle and becomes relatively less important when the derivative-controlled interlayer contributions grow at larger $\alpha$. This makes the first magic angle qualitatively distinct from the higher-order members of the chiral hierarchy~\cite{Navarro2022}.

We now consider the non-chiral off-diagonal contribution generated by the anticommutators in Eqs.~\eqref{eq:MD} and \eqref{eq:MDd}. The derivative part is
\begin{equation}
\begin{aligned}
\langle A_{g,s}\rangle&=i\int_{\rm u.c.}\left(-\chi_2^*\{B,p\}\psi_1+\chi_1^*\{B^*,p\}\psi_2\right)d^2\bm r\\
&=\sum_{\nu=1}^{3}\int_{\rm u.c.}\left(\chi_2^*\{e^{i\bm q_\nu\cdot\bm r},p\}\psi_1-\chi_1^*\{e^{-i\bm q_\nu\cdot\bm r},p\}\psi_2\right)d^2\bm r,
\end{aligned}
\end{equation}
where $p=p_x-ip_y$. This term is activated only when $\kappa\neq0$ and enters Eq.~\eqref{eq:decomp} with the prefactor $\kappa\alpha$. Because it also contains a momentum operator, it is strongly enhanced when the wave functions become more localized. This contribution is therefore one of the main energetic channels through which finite same-sublattice tunneling reorganizes the low-energy spectrum.

The anticommutator also contains a local derivative piece, since
\begin{equation}
\{B,p\}=2Bp+[p,B].
\end{equation}
The commutator term is proportional to a derivative of the pseudo-magnetic field and can be written as a local sublattice-mixing contribution. We denote its expectation value by
\begin{equation}
\langle A_{f,S}\rangle=\int_{\rm u.c.}\left(\chi_2^*\mathcal D_B\psi_1+\chi_1^*\mathcal D_B^*\psi_2\right)d^2\bm r,
\end{equation}
where $\mathcal D_B$ is the local operator obtained from the derivative of $B$. In the numerical results for the lowest-energy band considered here, this local sublattice-mixing contribution vanishes within numerical precision,
\begin{equation}
\langle A_{f,S}\rangle=0.
\end{equation}
Thus the dominant non-chiral inter-sublattice contribution comes from the derivative anticommutator term $\langle A_{g,s}\rangle$.

The scalar off-diagonal potential generated by the anticommutator is
\begin{equation}
V_S(\bm r)=U_0(\bm r)U_1(-\bm r)+U_0(-\bm r)U_1(\bm r),
\end{equation}
and its expectation value is
\begin{equation}
\langle V_S\rangle=\int_{\rm u.c.}\sum_j\chi_j^*\left[U_0(\bm r)U_1(-\bm r)+U_0(-\bm r)U_1(\bm r)\right]\psi_j\,d^2\bm r.
\end{equation}
For the $C_3$-symmetric states analyzed here, this expectation value vanishes by the symmetry of the moir\'e unit cell,
\begin{equation}
\langle V_S\rangle=0.
\end{equation}
Equivalently, the three first-star harmonic contributions transform into one another under $C_3$, and their phases cancel after integration over the unit cell. Therefore, although $V_S(\bm r)$ is an important operator in the squared Hamiltonian, its direct contribution to the total energy of these states vanishes. The non-chiral effects seen numerically are instead dominated by the confinement renormalization and by the derivative anticommutator channel $\langle A_{g,s}\rangle$.

Equation~\eqref{eq:decomp} therefore separates the physical mechanisms entering the non-chiral model. The term $\langle\bm A_f^2\rangle$ is the symmetric confinement energy. Near $\kappa\simeq0.7$, this term is close to the special uniform-confinement point $\kappa=1/\sqrt2$. The term $\langle\Delta\rangle$ gives the remaining antisymmetric pseudo-magnetic structure of the diagonal block. The term $\langle\bm A\cdot\nabla\rangle$ represents the chiral orbital interlayer contribution, while $\langle\bm B\cdot\bm\tau\rangle$ is the pseudo-Zeeman contribution. Finally, $\langle A_{g,s}\rangle$ is the leading non-chiral inter-sublattice contribution generated by finite same-sublattice tunneling.

In the exact chiral limit, $\kappa=0$, Eq.~\eqref{eq:decomp} reduces to the expectation-value equation of the TKV model,
\begin{equation}
\langle T\rangle+\alpha^2\langle\bm A^2\rangle-\alpha\left(\langle\bm A\cdot\nabla\rangle+\langle\bm B\cdot\bm\tau\rangle\right)=E^2.
\end{equation}
For the symmetry-adapted states considered here, the unit-cell average of the antisymmetric contribution does not contribute to the total energy,
\begin{equation}
\langle\Delta\rangle=0,
\end{equation}
so that the chiral energy balance is controlled by kinetic energy, symmetric confinement, and the orbital and pseudo-Zeeman interlayer terms.

\begin{figure}[t]
\centering
\includegraphics[width=\columnwidth]{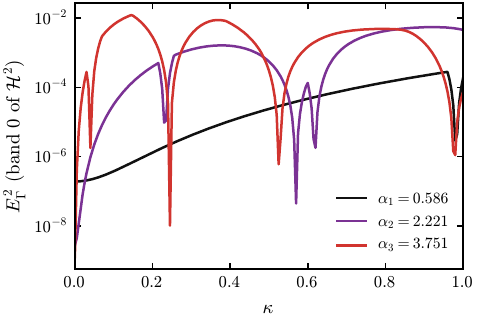}
\caption{Bandwidth $\mathrm{Log}(E_\Gamma)$ at the $\Gamma$ point in semi-logarithmic scale for the non-chiral model as a function of $\kappa$ for the first three magic angles, $\alpha_1=0.586$, $\alpha_2=2.221$, and $\alpha_3=3.751$. Band 0 refers to the lowest-energy band in $H^2$. Only the first magic angle shows a monotonic response to $\kappa$, while the higher-order magic angles exhibit oscillatory behavior.}
\label{fig:EGkappa}
\end{figure}

Figure~\ref{fig:E2} shows the behavior of $E^2(\alpha)$ at the $\Gamma$ point for both the chiral and relaxation-renormalized non-chiral models. In the exact chiral limit, the numerical data display the expected squeezing of the spectrum, with the band edge decreasing rapidly near the chiral magic angles. By contrast, in the non-chiral model with $\kappa\simeq0.7$, the squeezing mechanism is partially blocked. The expectation-value decomposition shows why this happens. Finite $\kappa$ places the diagonal confinement close to the uniform-confinement point, but it also activates the derivative anticommutator channel $\langle A_{g,s}\rangle$, which introduces an additional inter-sublattice energy scale. The realistic model therefore does not simply inherit the chiral cancellation mechanism; it replaces it by a new balance between confinement and competing interlayer channels.

More information is obtained from Fig.~\ref{fig:EGkappa}, where the low-energy scale is shown as a function of $\kappa$ for the first three chiral magic angles. The first magic angle behaves differently from the higher-order ones. For $\alpha_1=0.586$, the energy varies monotonically with $\kappa$, whereas higher-order magic angles display oscillatory behavior. This shows that the first magic angle is not merely the first member of the chiral sequence; it is the member that remains most compatible with the relaxation-renormalized energy balance of the non-chiral model.

\begin{figure}[t]
\centering
\includegraphics[width=\columnwidth]{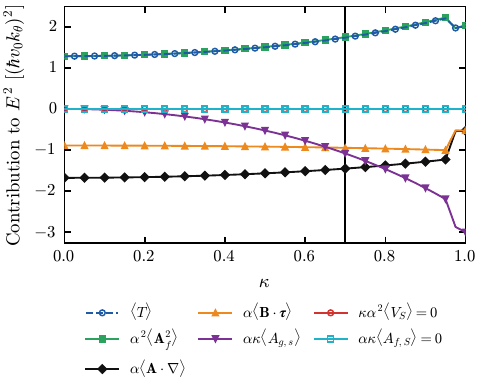}
\caption{Expected values for the Hamiltonian in Eq.~\eqref{eq:decomp} at the $\bm\Gamma$ point as a function of $\kappa$ for the first magic angle, $\alpha_1=0.586$. The vertical line indicates the experimentally relevant value of $\kappa$. Around $\kappa\simeq0.7$, the system lies close to the special confinement point $\kappa=1/\sqrt2$, where the oscillatory part of the symmetric confinement is strongly suppressed. In the same interval, the chiral and non-chiral interlayer contributions become comparable in magnitude. The contributions that vanish by symmetry or numerical cancellation are indicated in the labels.}
\label{fig:decompkappa}
\end{figure}

To further understand the role of lattice relaxation, Fig.~\ref{fig:decompkappa} shows the energetic contributions as a function of $\kappa$ for the first magic angle. The calculations are performed for the low-energy state associated with the flat-band sector at the $\Gamma$ point. The physically relevant interval $0.6\lesssim\kappa\lesssim0.8$ is precisely where the confinement channel approaches the special value $\kappa=1/\sqrt2$, and where the chiral and non-chiral interlayer contributions become comparable in magnitude. This is a central result of the expectation-value analysis: realistic relaxation places TBG close to a balanced regime in which the symmetric confinement is nearly uniform and the two interlayer channels compete on comparable energy scales.

\begin{figure}[t]
\centering
\includegraphics[width=\columnwidth]{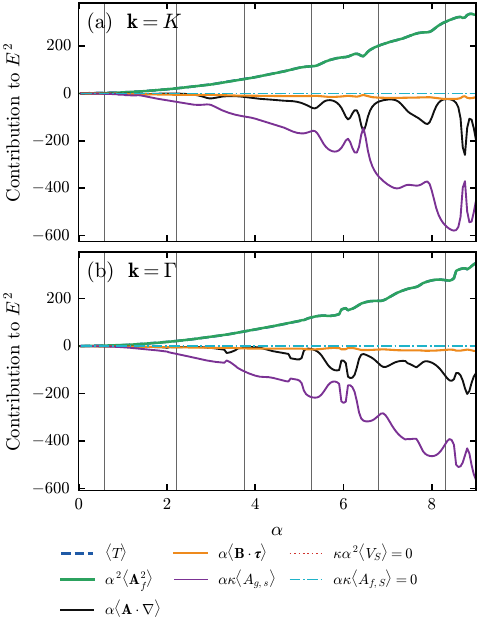}
\caption{(a) Expected values as a function of $\alpha$ for $\kappa=0.7$ at the Dirac point $\bm K$. (b) Expected values as a function of $\alpha$ for $\kappa=0.7$ at the $\bm\Gamma$ point. The plots reveal the interplay between confinement energy, kinetic energy, and the different interlayer contributions. The chiral and non-chiral channels exhibit an anticorrelated behavior, indicating an effective compensation between the orbital interlayer process and the relaxation-induced inter-sublattice process. The pseudo-Zeeman contribution is most relevant near the first magic angle. The contributions that vanish by symmetry or numerical cancellation are indicated in the labels.}
\label{fig:decompalpha}
\end{figure}

The dependence of the energetic contributions on the twist-angle parameter $\alpha$ is shown in Fig.~\ref{fig:decompalpha} for the realistic value $\kappa=0.7$. Panels (a) and (b) correspond to the high-symmetry points $\bm K$ and $\bm\Gamma$, respectively. The vertical black lines indicate the chiral-model magic angles.

Several important trends emerge from these plots. First, both the confinement energy and the kinetic energy increase systematically with $\alpha$, reflecting the stronger real-space localization of the electronic states at smaller twist angles. Second, the pseudo-Zeeman contribution is most relevant for $\alpha<1$, and therefore plays its most visible role near the first magic angle. Third, and most importantly, the chiral and non-chiral interlayer contributions display a pronounced complementary behavior. Whenever the chiral contribution $\langle\bm A\cdot\nabla\rangle$ develops a local minimum, the non-chiral contribution $\langle A_{g,s}\rangle$ tends to develop a corresponding maximum. This anticorrelation is the energetic signature of an effective compensation between the two interlayer channels. It does not imply an exact conservation law at the operator level, but it shows that the realistic non-chiral spectrum is governed by a strong redistribution of weight between the chiral orbital channel and the relaxation-induced inter-sublattice channel.

The expectation-value analysis therefore identifies the energetic origin of the robustness of the first magic angle beyond the chiral limit. Finite $\kappa$ brings the symmetric confinement close to a special nearly uniform regime, while simultaneously activating a non-chiral derivative channel that competes with the chiral orbital contribution. At the first magic angle these effects remain balanced, whereas at higher-order magic angles the balance becomes oscillatory and more strongly hybridized with remote bands. This motivates the current interpretation developed in the next section, where the two interlayer channels are analyzed directly.

\section{Currents between layers and sublattices}

The expectation-value analysis of the previous section shows that finite $\kappa$ activates an additional interlayer channel through the derivative anticommutator term $\langle A_{g,s}\rangle$. We now interpret this contribution in terms of current-like Fourier components of the squared Hamiltonian. As detailed in Ref.~\cite{Navarro2022}, the off-diagonal terms of the squared chiral Hamiltonian can be identified with the sum of the Fourier components of an interlayer current at the three first-star momenta $\bm q_1$, $\bm q_2$, and $\bm q_3$. In the present notation, the chiral orbital contribution is
\begin{equation}
\begin{aligned}
J_{AA}&\equiv\alpha\langle\bm A\cdot\nabla\rangle\\
&=2i\alpha\int_{\rm u.c.}\left(\psi_2^*\bm A\cdot\nabla\psi_1+\psi_1^*\bm A^*\cdot\nabla\psi_2\right)d^2\bm r.
\end{aligned}
\end{equation}
This quantity measures the current-like Fourier weight associated with the chiral interlayer channel. It is already present at $\kappa=0$, and it is controlled by the emergent gauge field $\bm A(\bm r)$ that appears in the squared chiral Hamiltonian.

Finite same-sublattice tunneling introduces a second current-like contribution through the non-chiral anticommutator term. We define
\begin{equation}
\begin{aligned}
J_{AB}&\equiv\alpha\kappa\langle A_{g,s}\rangle\\
&=i\alpha\kappa\int_{\rm u.c.}\left(-\chi_2^*\{B,p\}\psi_1+\chi_1^*\{B^*,p\}\psi_2\right)d^2\bm r.
\end{aligned}
\end{equation}
This term is absent in the chiral limit and is activated only when $\kappa\neq0$. It couples the two sublattice sectors $\Psi$ and $\chi$ through the derivative anticommutator $\{B,p\}$. For this reason, $J_{AB}$ represents a relaxation-induced inter-sublattice channel of the squared Hamiltonian.

In what follows, we refer to $J_{AA}$ and $J_{AB}$ as currents for brevity. More precisely, they are not macroscopic transport currents, but sums of current-like Fourier components at $\bm q_1$, $\bm q_2$, and $\bm q_3$. This distinction is important because the quantities are extracted from the squared Hamiltonian and measure how the off-diagonal operators redistribute spectral weight between layer and sublattice sectors.

The physical interpretation is direct. The chiral channel $J_{AA}$ is the orbital interlayer contribution inherited from the TKV limit. It preserves the chiral sector structure and participates in the cancellation mechanism responsible for spectral squeezing. The non-chiral channel $J_{AB}$, by contrast, is generated by finite same-sublattice tunneling in the original BM Hamiltonian and appears in the squared Hamiltonian as an inter-sublattice derivative channel. It therefore provides a leakage path between the two sublattice sectors. This leakage breaks the strict chiral cancellation pattern and is one of the mechanisms by which higher-order chiral magic angles are destabilized in the relaxation-renormalized model.

\begin{figure}[t]
\centering
\includegraphics[width=\columnwidth]{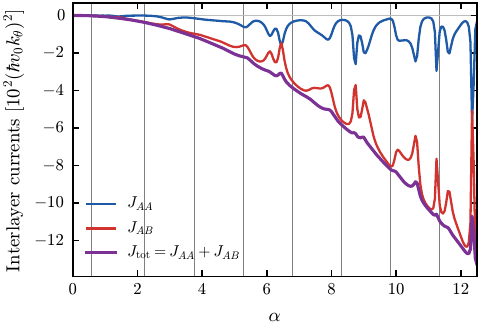}
\caption{Current-like interlayer Fourier components at the Dirac point $\bm K$ for the relaxation-renormalized value $\kappa\approx0.7$. The chiral orbital channel $J_{AA}$ is shown in blue, while the non-chiral inter-sublattice channel $J_{AB}$ is shown in red. The purple curve corresponds to the sum $J_{\rm tot}=J_{AA}+J_{AB}$. The strong oscillations of the individual channels are partially compensated in the total current-like contribution, revealing an effective current-balance relation in the non-chiral squared Hamiltonian.}
\label{fig:currents}
\end{figure}

To make this behavior explicit, Fig.~\ref{fig:currents} shows the two current-like contributions for the realistic non-chiral model at the Dirac point $\bm K$. Each contribution oscillates strongly as a function of $\alpha$, especially near the chiral magic angles. The important result is that these oscillations are not independent. The two channels display a pronounced complementary behavior: when the chiral contribution is enhanced, the non-chiral contribution is reduced, and vice versa. As a consequence, their sum remains much smoother than either component separately. The total current-like contribution,
\begin{equation}
J_{\rm tot}\equiv J_{AA}+J_{AB},
\end{equation}
is found numerically to follow the approximate scaling
\begin{equation}
J_{\rm tot}\approx-0.084\,\alpha^2.
\label{eq:Jtot}
\end{equation}
This empirical relation is one of the clearest signatures of the energy redistribution induced by finite $\kappa$. Although $J_{AA}$ and $J_{AB}$ separately show strong oscillatory structure, their sum tracks a smooth negative contribution with approximately quadratic scaling in $\alpha$.

This behavior can be understood from the energy balance in Eq.~\eqref{eq:expH2}. Near the magic-angle regime, the total squared energy $E^2$ remains small compared with the individual kinetic, confinement, and interlayer contributions. The diagonal terms of the squared Hamiltonian provide the positive kinetic and confinement channels, including the dominant $\alpha^2$ confinement scale. The off-diagonal current-like terms must therefore supply a compensating negative contribution. In the chiral model this compensation is carried only by the orbital channel $J_{AA}$. In the relaxation-renormalized model, the compensation is redistributed between $J_{AA}$ and the non-chiral inter-sublattice channel $J_{AB}$. This redistribution prevents the chiral cancellation pattern from remaining intact at higher-order magic angles and explains why the realistic spectrum deviates from the ideal TKV hierarchy.

The current-balance relation is therefore not an exact conservation law imposed at the operator level. Rather, it is an emergent energetic compensation visible in the expectation values of the low-energy states. Its importance is that it identifies how relaxation modifies the flat-band mechanism without completely destroying the chiral structure. At the first magic angle, the two current-like channels remain sufficiently balanced to preserve narrow central bands. At higher-order magic angles, the oscillatory redistribution becomes stronger, remote-band hybridization increases, and the exact chiral squeezing mechanism is lost.

\section{Breakdown of the 3/2 flux quantization rule}

In the chiral squared-Hamiltonian formulation, the higher-order magic angles can be organized in terms of an effective flux-quantization condition associated with the pseudo-magnetic confinement and current balance of the TKV limit. In this description, the chiral hierarchy follows the characteristic $3/2$ flux-scaling rule derived for the ideal model~\cite{Navarro2023,Navarro2022,Navarro2023RMF}. Finite same-sublattice tunneling modifies both the confinement landscape and the off-diagonal current-like channels, so the breakdown of the higher-order magic-angle sequence can be viewed as the breakdown of this chiral $3/2$ quantization structure.

The previous sections showed that the non-chiral model remains continuously connected to the chiral TKV limit, but with important corrections generated by finite same-sublattice tunneling. An important consequence is that the first magic angle retains many of the characteristic features of the chiral model, whereas higher-order magic angles are much more strongly modified. This robustness is particularly significant because the strongest correlated and superconducting phenomena in twisted bilayer graphene are observed precisely near the first magic angle. A central question, therefore, is why the first magic angle is so special.

Part of the answer follows from the perturbative analysis. As discussed in the Appendix, the leading non-chiral corrections at small $\alpha$ scale as $(\kappa\alpha)^2$. This shows that the spectrum remains comparatively close to the chiral limit as the first magic-angle regime is approached from small $\alpha$. At the first magic angle itself, $\alpha_1\approx0.586$, the perturbative expansion is no longer quantitatively controlled, and the persistence of the narrow-band structure must instead be understood from the non-perturbative confinement and current-channel balance discussed above. For higher-order magic angles, the larger values of $\alpha$ further amplify the non-chiral contributions and remote-band hybridization. Thus, the perturbative result explains the initial robustness of the chiral structure, while the current-channel analysis identifies the mechanism that sustains the first magic angle beyond the perturbative regime.

\begin{figure*}[t]
\centering
\includegraphics[width=\textwidth]{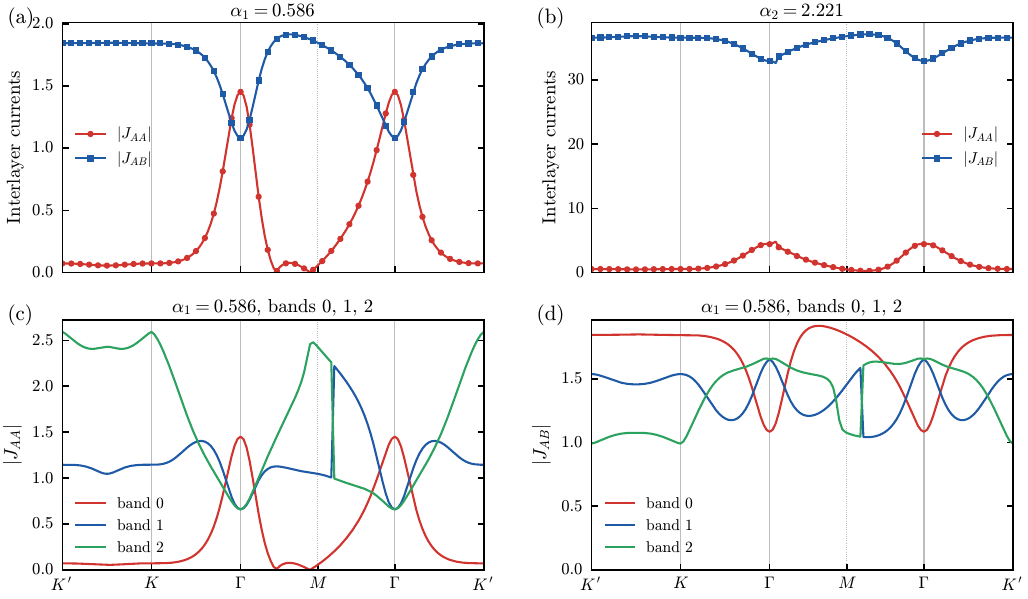}
\caption{Interlayer Fourier components of the current-like contributions in the non-chiral model for $\kappa\approx0.7$. (a) $|J_{AA}|$ (red) and $|J_{AB}|$ (blue) at the first magic angle $\alpha_1\approx0.586$. (b) Same quantities for the second magic angle $\alpha_2\approx2.221$. The overlap between both channels near the $\Gamma$ point is a distinctive feature of the first magic angle and disappears at higher-order magic angles. Panels (c) and (d) show the contributions of the three lowest bands to $J_{AA}$ and $J_{AB}$, respectively, at the first magic angle. Higher remote bands contribute more strongly to the non-chiral channel.}
\label{fig:currentsk}
\end{figure*}

Figure~\ref{fig:currentsk} provides additional information. Panels (a) and (b) show the two current-like contributions defined in the previous section, $J_{AA}$ and $J_{AB}$, as functions of momentum for the first and second magic angles, respectively. At the first magic angle, $\alpha_1\approx0.586$, the two channels become comparable near the $\Gamma$ point.

This overlap is absent at the second magic angle, $\alpha_2\approx2.221$, where the two channels are clearly separated. Thus, the first magic angle corresponds to a special balance point between the chiral orbital current channel and the relaxation-induced inter-sublattice current channel. This balance is lost as one moves to higher-order magic angles.

This result suggests a useful analogy with heavy-fermion systems, where an itinerant channel hybridizes with a more localized one. In the present case, $J_{AA}$ plays the role of the itinerant current channel inherited from the chiral model, whereas $J_{AB}$ behaves as a more localized, relaxation-induced channel that is activated only in the non-chiral model. The overlap between both channels near the $\Gamma$ point at the first magic angle suggests a natural mechanism for the formation of strongly correlated narrow bands. In this sense, the relaxation-induced channel can be viewed as an effective localized component embedded within the itinerant topological flat-band structure, and the overlap between both channels near high-symmetry points is reminiscent of a Kondo-like hybridization mechanism~\cite{Song2022b,Calugaru2023}. We stress that this is an analogy at the level of channel hybridization, but it provides a suggestive framework for interpreting why the first magic angle is particularly favorable for correlated phenomena.

The role of higher-energy bands is also crucial. Panels (c) and (d) of Fig.~\ref{fig:currentsk} show the contributions of the three lowest bands to $J_{AA}$ and $J_{AB}$ at the first magic angle. For the chiral current $J_{AA}$, the flat band dominates near the $\Gamma$ point, while the higher-energy bands give only relatively small corrections. By contrast, the non-chiral channel $J_{AB}$ receives substantially stronger contributions from the remote bands. This shows that remote-band hybridization selectively enhances the non-chiral inter-sublattice channel.

This distinction is important for understanding the breakdown of the $3/2$ flux quantization rule. In the ideal chiral theory, the flat-band condition is protected by a delicate balance encoded in the chiral current channel. Once finite $\kappa$ is introduced, the additional non-chiral channel $J_{AB}$ grows through hybridization with remote bands and progressively disrupts that balance. The first magic angle remains special because the overlap between $J_{AA}$ and $J_{AB}$ near $\Gamma$ still allows a balanced low-energy regime. At higher-order magic angles, however, this overlap is lost and the non-chiral channel becomes increasingly dominated by remote-band processes. The resulting imbalance provides a direct microscopic mechanism for the breakdown of the chiral $3/2$ flux quantization rule in realistic non-chiral twisted bilayer graphene.

\section{Conclusions}

In this work, we analyzed why the first magic angle of twisted bilayer graphene remains robust beyond the ideal chiral limit, while higher-order magic angles are strongly destabilized by realistic same-sublattice tunneling. We worked within the non-chiral Bistritzer--MacDonald continuum Hamiltonian, where lattice relaxation is incorporated through the finite tunneling ratio $\kappa=w_{AA}/w_{AB}$. By extending the squared-Hamiltonian formulation previously developed for the chiral model, we obtained an effective description in which the role of finite $\kappa$ can be separated into diagonal confinement renormalization and off-diagonal inter-sublattice coupling terms.

A central result is that finite $\kappa$ does not simply perturb the chiral flat-band problem. Instead, it reorganizes the effective confinement landscape of the squared Hamiltonian. The diagonal confinement potential is renormalized according to
\begin{equation*}
\bm A^2\to\bm A_f^2,
\end{equation*}
with an oscillatory component proportional to $-1+2\kappa^2$. Therefore, the special value
\begin{equation*}
\kappa=\frac{1}{\sqrt2}\simeq0.707
\end{equation*}
cancels the spatial modulation of the symmetric confinement term. This value is strikingly close to the relaxation-renormalized value $\kappa\simeq0.7$ commonly used for realistic TBG. Thus, realistic lattice relaxation places the continuum model close to a special confinement point of the squared Hamiltonian, where the symmetric confinement becomes nearly uniform and the remaining diagonal spatial structure is controlled mainly by the antisymmetric pseudo-magnetic term. This provides a direct analytical interpretation of the relaxation-renormalized BM parameter and suggests that lattice relaxation favors a nearly optimized confinement regime.

This confinement mechanism has clear spectral and wave-function consequences. In the chiral limit, the magic-angle hierarchy is governed by an exact scaling structure that produces ideal spectral squeezing and exactly flat bands at the TKV magic angles. Once finite $\kappa$ is included, this scaling structure is partially broken. The first magic angle remains narrow and continuously connected to the chiral flat-band regime, but the higher-order magic angles develop substantial residual dispersion and stronger hybridization with remote bands. The loss of ideal squeezing is therefore not a numerical accident; it is the spectral consequence of the finite-$\kappa$ terms that compete with the chiral cancellation mechanism.

The real-space and Fourier-space analyses support the same conclusion. Finite $\kappa$ squeezes the low-energy wave functions around the AA regions. In reciprocal space, this appears as a shift of the dominant Fourier components toward larger moir\'e momenta, especially at higher-order magic angles. Therefore, lattice relaxation simultaneously enhances real-space localization and increases coupling to remote bands. This dual effect explains why isolated-flat-band descriptions remain most reliable near the first magic angle but become progressively less controlled for higher-order chiral magic angles.

The expectation-value analysis identified the energetic channels responsible for this behavior. The kinetic energy, the renormalized confinement energy, the antisymmetric pseudo-magnetic contribution, and the off-diagonal interlayer terms all contribute to the balance of the squared energy. Near $\kappa\simeq0.7$, the system lies close to the special uniform-confinement point, while the chiral and non-chiral interlayer contributions become comparable in magnitude. This is a key result: realistic TBG is not merely a weakly broken version of the chiral model, but a system in which the relaxation-renormalized confinement and the interlayer current-like channels enter a balanced regime.

In the squared-Hamiltonian language, the two relevant interlayer channels are the chiral orbital channel $J_{AA}$ and the relaxation-induced inter-sublattice channel $J_{AB}$. The first is inherited from the chiral TKV model, while the second is activated only by finite $\kappa$. These two contributions exhibit a pronounced complementary behavior: their individual values oscillate strongly as functions of $\alpha$, but their sum remains comparatively smooth and follows an approximate quadratic scaling. This current-balance relation is not an exact operator conservation law, but it is an emergent energetic compensation in the low-energy states. It reveals how the non-chiral model redistributes the current-like spectral weight that, in the chiral limit, is carried only by the orbital channel.

This redistribution also clarifies the breakdown of the $3/2$ flux-quantization rule at higher-order magic angles. At the first magic angle, the two current-like channels overlap near the $\Gamma$ point, producing a balanced low-energy regime. At the second and higher magic angles, this overlap is lost. Moreover, remote bands contribute more strongly to the non-chiral channel $J_{AB}$ than to the chiral channel $J_{AA}$. As a result, hybridization with remote bands selectively enhances the relaxation-induced inter-sublattice channel and destabilizes the delicate chiral balance required for the higher-order magic-angle hierarchy.

The main physical picture that emerges is therefore the following. Lattice relaxation drives realistic TBG close to a special confinement regime of the squared Hamiltonian, while finite same-sublattice tunneling activates an additional inter-sublattice channel that competes with the chiral orbital current. At the first magic angle these effects remain balanced, allowing narrow central bands and preserving important remnants of the chiral theory. At higher-order magic angles the balance becomes unstable: remote-band hybridization grows, the current-channel compensation becomes oscillatory, and the exact chiral squeezing mechanism is lost.

The squared-Hamiltonian framework developed here provides a compact way to understand these effects from a single-particle perspective. It connects lattice relaxation, pseudo-magnetic confinement, real-space localization, remote-band hybridization, and interlayer current redistribution within one effective description. More broadly, this approach suggests that realistic moir\'e systems beyond the chiral approximation can be understood in terms of emergent gauge-field structures and symmetry-constrained current-like channels, rather than only through numerical band-structure deformation.

\begin{acknowledgments}
We thank Danna Liu for useful discussions. This work was supported by CONAHCyT project 1564464 and UNAM DGAPA project IN101924 and by the National Natural Science Foundation of China (NSFC) Grant.\ No.\ 125471103. L.A.N.L.\ was supported by a CONAHCyT PhD scholarship No.\ 1564464. F.G.\ acknowledge funding from the EU NextGenerationEU/PRTR-C17.I1 and the IKUR Strategy under the collaboration agreement between Ikerbasque Foundation and DIPC on behalf of the Department of Education of the Basque Government. P.A.P and F.G acknowledges support from the ``Severo Ochoa'' Programme for Centres of Excellence in R\&D (CEX2020-001039-S/AEI/10.13039/501100011033), NOVMOMAT Grant PID2022-142162NB-I00 funded by MCIN/AEI/10.13039/501100011033 and by ``ERDF A way of making Europe'', P.A.P.\ acknowledges support Grant No.\ JSF-24-05-0002 of the Julian Schwinger Foundation for Physics Research.
\end{acknowledgments}

\section*{Appendix A: Perturbative treatment of the non-chiral term}
\label{app:pert}

In this appendix we derive the leading analytical correction produced by the finite same-sublattice tunneling term in the small-$\alpha$ regime. The purpose is twofold. First, the calculation verifies the numerical results at small twist-angle coupling. Second, it explains why the first magic angle remains comparatively robust beyond the chiral limit: the projected first-order correction vanishes, and the leading non-chiral shift appears only at order $\kappa^2\alpha^2$.

We decompose the non-chiral BM Hamiltonian as
\begin{equation}
\mathcal H=H_D+H_M,
\end{equation}
with
\begin{equation}
H_D=\begin{pmatrix}0 & D^\dagger\\ D & 0\end{pmatrix},\qquad
H_M=\begin{pmatrix}M & 0\\ 0 & M\end{pmatrix}.
\end{equation}
Here $H_D$ is the chiral BM Hamiltonian, while $H_M$ contains the finite same-sublattice tunneling generated by lattice relaxation. The block $M$ is the same operator defined in Eq.~\eqref{eq:M}. In the notation of the main text, it is diagonal in the $(\Psi,\chi)$ sublattice-block structure and off diagonal in layer space,
\begin{equation}
M(\bm r)=\kappa\alpha\begin{pmatrix}0 & U_0(\bm r)\\ U_0(-\bm r) & 0\end{pmatrix}.
\end{equation}
Therefore $H_M$ scales as $\kappa\alpha$.

\begin{figure}[t]
\centering
\includegraphics[width=\columnwidth]{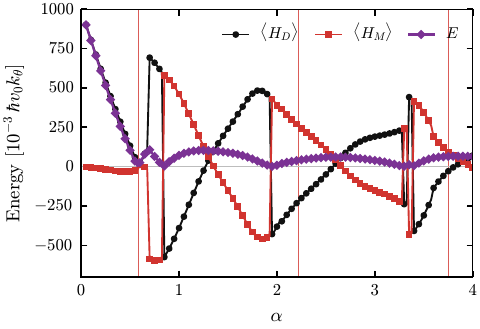}
\caption{Expected values of the operators $H_D$ and $H_M$ as functions of the twist-angle coupling $\alpha$, evaluated using the exact numerical state of the central band at the $\Gamma$ point for $\kappa=0.7$. The total energy of the central band is shown as a reference. The vertical red lines indicate the chiral magic angles. The comparison shows that the perturbative treatment is controlled only in the small-$\alpha$ regime, while for larger $\alpha$ the chiral and non-chiral contributions become strongly intertwined.}
\label{fig:HDHM}
\end{figure}

\begin{figure*}[t]
\centering
\includegraphics[width=\textwidth]{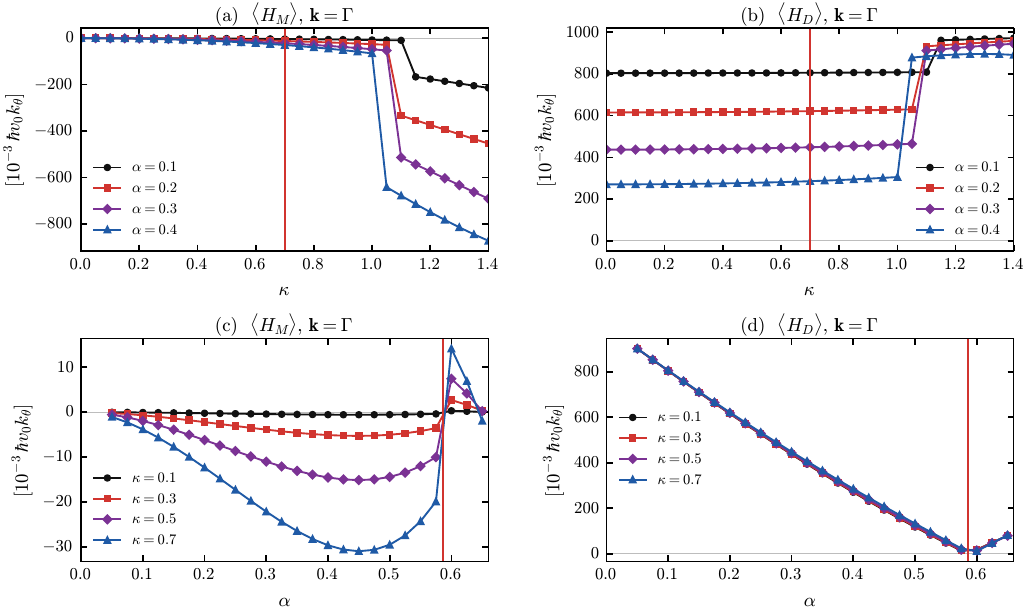}
\caption{Expected values of $H_M$ and $H_D$ using the exact numerical state of the central band at the $\Gamma$ point. The evolution of both contributions illustrates the crossover from the perturbative regime at small $\alpha$ to the strongly mixed regime near and beyond the first magic angle.}
\label{fig:HDHMpanels}
\end{figure*}

At the $\Gamma$ point, $\bm k=\bm q_1$, the chiral zero-mode wave function used in the perturbative calculation can be written as
\begin{equation}
|\Phi^{(0)}_{\mu_\alpha}\rangle=\mathcal N_0\begin{pmatrix}\psi_1^{(0)}(\bm r)\\ \psi_2^{(0)}(\bm r)\\ 0\\ 0\end{pmatrix},
\label{eq:Phi0}
\end{equation}
where
\begin{equation}
\begin{aligned}
\psi_1^{(0)}(\bm r)&=U_1(-\bm r)+\frac{\alpha}{3}U_1(2\bm r),\\
\psi_2^{(0)}(\bm r)&=i\mu_\alpha\left[U_1(\bm r)+\frac{\alpha}{3}U_1(-2\bm r)\right].
\end{aligned}
\label{eq:psi0}
\end{equation}
The label $\mu_\alpha=\pm1$ distinguishes the two symmetry-related zero-mode branches, and
\begin{equation}
\mathcal N_0=\left[2\left(3+\frac{\alpha^2}{3}\right)\right]^{-1/2}
\end{equation}
is the normalization factor in the first-star approximation. Figures~\ref{fig:HDHM} and \ref{fig:HDHMpanels} show that this perturbative description is reliable only at small $\alpha$. Near higher-order magic angles, the expectation values of $H_D$ and $H_M$ become strongly mixed, consistent with the breakdown of the simple chiral scaling picture discussed in the main text.

\subsection*{A.1 Vanishing of the first-order correction}

The first-order correction is the matrix element of $H_M$ projected onto the unperturbed chiral zero-mode subspace,
\begin{equation}
E^{(1)}=\langle\Phi^{(0)}_{\mu_\alpha}|H_M|\Phi^{(0)}_{\mu_\alpha}\rangle.
\end{equation}
Using Eqs.~\eqref{eq:Phi0} and \eqref{eq:psi0}, this becomes
\begin{multline}
E^{(1)}=\kappa\alpha\mathcal N_0^2\int_{\rm u.c.}d^2\bm r\Big[\big(\psi_1^{(0)}\big)^*U_0(\bm r)\psi_2^{(0)}\\
+\big(\psi_2^{(0)}\big)^*U_0(-\bm r)\psi_1^{(0)}\Big].
\end{multline}
Substituting the explicit zero-mode components gives, to leading order in the first-star expansion,
\begin{multline}
E^{(1)}=i\mu_\alpha\kappa\alpha\mathcal N_0^2\int_{\rm u.c.}d^2\bm r\Big[U_1^\dagger(-\bm r)U_0(\bm r)U_1(\bm r)\\
-U_1^\dagger(\bm r)U_0(-\bm r)U_1(-\bm r)\Big]+\mathcal O(\kappa\alpha^2).
\label{eq:E1}
\end{multline}

The unit-cell integral of a product of moir\'e harmonics is nonzero only if the total momentum carried by the product is a reciprocal lattice vector of the moir\'e superlattice. Equivalently,
\begin{equation}
\int_{\rm u.c.}e^{i(\bm k_a+\bm k_b+\bm k_c)\cdot\bm r}\,d^2\bm r\neq0
\end{equation}
only when
\begin{equation}
\bm k_a+\bm k_b+\bm k_c=\bm G.
\end{equation}
For the $\Gamma$-point chiral zero mode, the products appearing in Eq.~\eqref{eq:E1} do not contain a $C_3$-invariant zero-momentum component. The three first-star contributions are cyclically related by $C_3$, and their phases cancel in the unit-cell average. Thus the projection of $H_M$ onto the chiral zero-mode subspace vanishes,
\begin{equation}
E^{(1)}=0.
\label{eq:E1zero}
\end{equation}
This is the key perturbative result. Finite same-sublattice tunneling is present already at order $\kappa\alpha$ in the Hamiltonian, but its first-order energy correction to the $\Gamma$-point chiral zero mode vanishes by the momentum-selection and $C_3$-symmetry structure of the unperturbed state.

\subsection*{A.2 Leading second-order correction}

Because the first-order correction vanishes, the leading non-chiral shift is second order in $H_M$. Standard Schr\"odinger perturbation theory gives
\begin{equation}
E^{(2)}=\sum_{n\neq0}\frac{\left|\langle\Phi_n^{(0)}|H_M|\Phi^{(0)}_{\mu_\alpha}\rangle\right|^2}{E_0^{(0)}-E_n^{(0)}},
\label{eq:E2PT}
\end{equation}
where $|\Phi_n^{(0)}\rangle$ are eigenstates of the chiral Hamiltonian $H_D$. Since $H_M\propto\kappa\alpha$, Eq.~\eqref{eq:E2PT} immediately implies the leading scaling
\begin{equation}
E^{(2)}\sim\kappa^2\alpha^2.
\end{equation}
This scaling is the perturbative origin of the robustness of the first magic angle: for $\alpha\lesssim\alpha_1$, the finite-$\kappa$ correction remains controlled, whereas for higher-order magic angles the same correction is amplified by the larger value of $\alpha$ and by stronger remote-band hybridization.

To obtain an explicit analytical approximation, we introduce the projector $\mathcal Q$ onto the remote-band subspace,
\begin{equation}
\mathcal Q=1-|\Phi^{(0)}_{\mu_\alpha}\rangle\langle\Phi^{(0)}_{\mu_\alpha}|,
\end{equation}
and replace the weighted remote-band denominators by an effective gap $\Delta E_{\rm eff}$. This gives the closure approximation
\begin{equation}
\begin{aligned}
E^{(2)}&\approx-\frac{1}{\Delta E_{\rm eff}}\langle\Phi^{(0)}_{\mu_\alpha}|H_M\mathcal QH_M|\Phi^{(0)}_{\mu_\alpha}\rangle\\
&=-\frac{1}{\Delta E_{\rm eff}}\left[\langle\Phi^{(0)}_{\mu_\alpha}|H_M^2|\Phi^{(0)}_{\mu_\alpha}\rangle-\big(E^{(1)}\big)^2\right].
\end{aligned}
\end{equation}

\begin{figure}[t]
\centering
\includegraphics[width=\columnwidth]{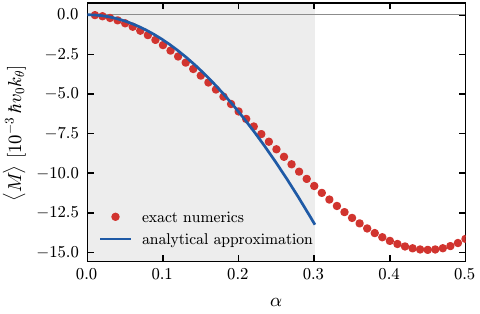}
\caption{Second-order energy shift $\langle M\rangle$ as a function of $\alpha$ for $\kappa=0.7$ at the $\bm k=\bm\Gamma$ point. Red dots correspond to exact numerical calculations, while the blue solid curve is the analytical approximation in Eq.~\eqref{eq:Mfinal}. The agreement is excellent in the controlled perturbative regime $\alpha\lesssim0.3$.}
\label{fig:pert}
\end{figure}

Using Eq.~\eqref{eq:E1zero}, this reduces to
\begin{equation}
E^{(2)}\approx-\frac{1}{\Delta E_{\rm eff}}\langle\Phi^{(0)}_{\mu_\alpha}|H_M^2|\Phi^{(0)}_{\mu_\alpha}\rangle.
\label{eq:E2closure}
\end{equation}
In the dimensionless BM units used throughout the paper, the leading remote-band denominators at small $\alpha$ are of order unity. Therefore the dominant dependence of Eq.~\eqref{eq:E2closure} is controlled by the matrix element of $H_M^2$, and the leading scaling remains $\kappa^2\alpha^2$.

The square of the relaxation block is
\begin{equation}
M^2=\alpha^2\kappa^2\begin{pmatrix}U_0(\bm r)U_0(-\bm r) & 0\\ 0 & U_0(-\bm r)U_0(\bm r)\end{pmatrix}.
\end{equation}
Using $U_0(-\bm r)=U_0^\dagger(\bm r)$, the diagonal product defines the scalar moir\'e field
\begin{equation}
\mathcal V_0(\bm r)=U_0(\bm r)U_0(-\bm r)=|U_0(\bm r)|^2.
\end{equation}
Thus
\begin{multline}
\langle\Phi^{(0)}_{\mu_\alpha}|H_M^2|\Phi^{(0)}_{\mu_\alpha}\rangle\\
=\alpha^2\kappa^2\mathcal N_0^2\int_{\rm u.c.}d^2\bm r\,\mathcal V_0(\bm r)\left[|\psi_1^{(0)}(\bm r)|^2+|\psi_2^{(0)}(\bm r)|^2\right].
\end{multline}

The density of the unperturbed state contains the harmonic expansion
\begin{equation}
\begin{aligned}
|\psi_1^{(0)}(\bm r)|^2={}&|U_1(-\bm r)|^2\\
&+\frac{\alpha}{3}\left[U_1^\dagger(-\bm r)U_1(2\bm r)+\text{h.c.}\right]\\
&+\frac{\alpha^2}{9}|U_1(2\bm r)|^2,
\end{aligned}
\end{equation}
with an analogous expression for $|\psi_2^{(0)}|^2$. After integration over the moir\'e unit cell, only the zero-momentum harmonic components survive. The resulting overlap coefficients determine the analytical numerator of the perturbative correction. In the gauge and first-star convention used here, the explicit harmonic evaluation gives the compact structure
\begin{equation}
E^{(2)}\propto\kappa^2\mu_\alpha\left(-2\alpha^2+\frac{\alpha^3}{9}\right),
\label{eq:E2num}
\end{equation}
where the $\alpha^2$ term is the leading non-chiral correction and the $\alpha^3$ term is the first higher-harmonic correction within the same expansion. The factor $\mu_\alpha=\pm1$ encodes the branch dependence of the $\Gamma$-point chiral zero mode.

To compare directly with the exact numerical data, we keep the leading normalization and overlap corrections generated by the same harmonic expansion. This gives the effective normalization factor
\begin{equation}
\mathcal N_{\rm eff}(\alpha,\kappa)=2\left[3+0.765\,\kappa\alpha+\frac{\alpha^2}{3}\left(1-1.53\,\kappa+9\kappa^2\right)\right].
\label{eq:Neff}
\end{equation}
The numerical coefficients in Eq.~\eqref{eq:Neff} are fixed by the unit-cell harmonic overlaps of the $U_m$ functions in the convention of Eq.~\eqref{eq:psi0}; they are not fitting parameters. Combining Eqs.~\eqref{eq:E2num} and \eqref{eq:Neff} gives the analytical approximation
\begin{equation}
\langle M\rangle=E^{(2)}\approx\frac{\kappa^2\mu_\alpha\left(-2\alpha^2+\frac{\alpha^3}{9}\right)}{2\left[3+0.765\,\kappa\alpha+\frac{\alpha^2}{3}\left(1-1.53\,\kappa+9\kappa^2\right)\right]}.
\label{eq:Mfinal}
\end{equation}
This expression makes explicit that the leading non-chiral correction is controlled by $\kappa^2\alpha^2$, with higher-order corrections entering at larger $\alpha$.

Figure~\ref{fig:pert} compares Eq.~\eqref{eq:Mfinal} with the numerical calculation. The agreement is excellent for $\alpha\lesssim0.3$, confirming the validity of the perturbative expansion in the small-$\alpha$ regime. Deviations at larger $\alpha$ are expected and physically meaningful: higher harmonics, remote-band hybridization, and the non-chiral current channel become increasingly important as the system approaches the first magic angle and beyond.

The important conclusion is that the finite-$\kappa$ correction is not linear in $\kappa\alpha$ at the $\Gamma$-point chiral zero mode. The first-order term vanishes by symmetry, and the leading correction enters as
\begin{equation}
E^{(2)}\sim\kappa^2\alpha^2.
\end{equation}
This result explains why the first magic angle preserves several features of the chiral model despite the realistic value $\kappa\simeq0.7$. At the same time, it also explains why higher-order magic angles are much less robust: increasing $\alpha$ pushes the system out of the perturbative regime, where confinement reshaping, current-channel redistribution, and remote-band hybridization dominate the spectrum.

\end{document}